\documentclass[transmag]{IEEEtran}
\usepackage{latexsym}
\usepackage{graphicx}
\usepackage{amsfonts,amssymb,amsmath}
\usepackage{hyperref}
\usepackage{cite}
\usepackage{booktabs}
\usepackage{bbold}
\usepackage{tipa}
\usepackage{scalerel}
\usepackage{xcolor}
\usepackage{url}
\usepackage{amssymb}
\usepackage{stackengine}
\DeclareMathOperator*{\argmax}{argmax} 
\newcommand{\beginsupplement}{%
        \setcounter{table}{0}
        \setcounter{page}{1}
        \setcounter{equation}{0}
        \setcounter{section}{0}
        \renewcommand{\thetable}{S\arabic{table}}%
        \setcounter{figure}{0}
        \renewcommand{\thefigure}{S\arabic{figure}}%
     }
\usepackage{amsmath,bm}
\usepackage{pifont}
\usepackage{multirow}
\usepackage{subcaption}
\def\BibTeX{{\rm B\kern-.05em{\sc i\kern-.025em b}\kern-.08em T\kern-.1667em\lower.7ex\hbox{E}\kern-.125emX}}
\begin{document}

\title{The effectiveness of unsupervised subword modeling with autoregressive and cross-lingual phone-aware networks}
\author{Siyuan Feng and Odette Scharenborg,~\IEEEmembership{Senior~Member,~IEEE}%
\thanks{The authors are with Multimedia Computing Group, Delft University of Technology, Delft, the Netherlands (e-mail: S.Feng@tudelft.nl; O.E.Scharenborg@tudelft.nl).}
\thanks{This manuscript has been accepted for publication in IEEE Open Journal of Signal Processing (OJ-SP). This pre-print is not the fully edited version.}}

\IEEEtitleabstractindextext{\begin{abstract}
This study addresses unsupervised subword modeling, i.e., learning acoustic feature representations that can distinguish between subword units of a language. We propose a two-stage learning framework that combines self-supervised learning and cross-lingual knowledge transfer. The framework consists of autoregressive predictive coding (APC) as the front-end and a cross-lingual deep neural network (DNN) as the back-end. 
Experiments on the ABX subword discriminability task conducted with the Libri-light and ZeroSpeech 2017 databases showed that our approach is competitive or superior to state-of-the-art studies. 
Comprehensive and systematic analyses at the phoneme- and articulatory feature (AF)-level showed that our approach was better at capturing diphthong than monophthong vowel information, while also differences in the amount of information captured for different types of consonants were observed. Moreover, a positive correlation was found between the effectiveness of the back-end in capturing a phoneme's information and the quality of the cross-lingual phone labels assigned to the phoneme. The AF-level analysis together with t-SNE visualization results showed that the proposed approach is better than MFCC and APC features in capturing manner and place of articulation information, vowel height, and backness information. 
Taken together, the analyses showed that the two stages in our approach are both effective in capturing phoneme and AF information. Nevertheless, monophthong vowel information is less well captured than consonant information, which suggests that future research should focus on improving capturing monophthong vowel information.

\end{abstract}


}

\maketitle

\section{INTRODUCTION}

There are around $7,000$ spoken languages in the world \cite{austin2011cambridge}.  For most of them, the amount of transcribed speech data resources is very limited, or even non-existent \cite{speech2020scharenborg}. Many of these low-resource languages, such as ethnic minority languages in China and languages in Africa, may have never been formally studied. In addition to the lack of enough transcribed speech data, linguistic knowledge about such languages is incomplete, or may even be entirely lacking. Conventional supervised acoustic modeling \cite{hinton2012deep,dahl2011context} can therefore not be applied directly. This leads to the current situation that high-performance ASR systems are only available for a small number of major languages, e.g., English, Mandarin, French. To facilitate ASR technology for low-resource languages, investigation of unsupervised acoustic modeling (UAM) methods is necessary, which aims to find and model a set of basic speech units that represents all the sounds in the language of interest, i.e., the low-resource, target language.

Recently, there has been a growing research interest in UAM \cite{lee2012a,I3EWang,chen2015parallel,ondel2017bayesian,Tjandra2019,Feng2019combining}. 
A strict assumption of UAM is that for the target language only raw speech data is available, while the transcriptions, phoneme inventory (and its size) and pronunciation lexicon are unknown. This is known as the \textit{zero-resource} assumption \cite{versteegh2015zero}.  
There are two main research strands in UAM. The first strand formulates the problem as discovering a finite set of phoneme-like speech units \cite{lee2012a,I3EWang,lee2015unsupervised,ondel2018bayesian}. This is often referred to as \textit{acoustic unit/model discovery} (AUD) \cite{lee2012a,ondel2017bayesian}. The second strand formulates the problem as learning acoustic feature representations that can distinguish subword (phoneme) units of the target language, and is robust to linguistically-irrelevant factors, such as speaker  \cite{heck2017feature,oord2017neural,feng2019_TASLP}. This is often referred to as \textit{unsupervised subword modeling} \cite{versteegh2015zero,dunbar2017zero,feng2019_TASLP}. In essence, the second strand is focused on learning an intermediate representation towards the ultimate goal of UAM, while the first strand aims directly at the ultimate goal. These two strands are closely connected and can benefit from each other; for instance, a good
subword-discriminative feature representation 
has been shown beneficial to AUD \cite{Chen+2016,feng2016exploit}, while conversely,  discovered speech units with good consistency with true phonemes are helpful to
learning   subword-discriminative acoustic feature representations \cite{heck2017feature,feng2019_TASLP}. 

This study addresses unsupervised subword modeling in UAM. Learning subword-discriminative feature representations in the zero-resource scenario has been shown to be a non-trivial task \cite{versteegh2015zero,dunbar2017zero}. The major difficulty is the separation of linguistic information (e.g., phoneme information)  from non-linguistic information (e.g., speaker information).  
For instance, a speech sound such as [\ae]\footnote{International Phonetic Alphabet (IPA) symbol.} produced by different speakers  might be mistakenly modeled as different speech units \cite{lee2015spoken}.

There are many interesting attempts to unsupervised subword modeling \cite{chen2015parallel,heck2017feature,oord2017neural,feng2019_TASLP,Tjandra2019,kahn2019librilight}. One typical research direction is to leverage purely unsupervised learning techniques. One method is the clustering of speech sounds that have acoustically similar patterns and that potentially correspond to the same subword units  \cite{chen2015parallel,ansari2017deep}, which results in phoneme-like pseudo transcriptions that can be used to facilitate subword-discriminative feature learning \cite{chen2015parallel,heck2017feature}.
Unsupervised and self-supervised representation learning algorithms are applied to learn, without using external supervision, speech features that retain the linguistic content in the original data while ignoring linguistically-irrelevant information, particularly speaker variation  \cite{oord2018cpc,oord2017neural,hsu2017nips,last2020unsupervised,Zeghidour+2016}.

A second research direction to unsupervised subword modeling is to exploit cross-lingual knowledge \cite{shibata2017composite,feng2018exploiting}. Speech and text resources from out-of-domain (OOD) resource-rich languages have been shown beneficial to modeling subword units of in-domain low-resource languages. 
\textcolor{black}{For instance, \cite{shibata2017composite,feng2018exploiting} used an OOD AM to extract cross-lingual bottleneck features (BNFs), and \cite{feng2018exploiting} also used an OOD ASR to generate cross-lingual phone labels.}


This study adopts a two-stage learning framework which combines both research directions within the area of unsupervised subword modeling \textcolor{black}{(the second research strand in UAM)}. 
At the first stage, the front-end, a self-supervised representation learning model named autoregressive predictive coding (APC)   \cite{Chung2019}  is trained. APC preserves phonetic (subword) and speaker information from the original speech signal, but makes the two information types more separable \cite{Chung2019}. 

At the second stage, the back-end, a cross-lingual, OOD DNN model with a bottleneck layer (DNN-BNF) is trained using the APC pretrained features as the input features to create the missing (due to the zero-resource assumption) frame labels.
This system framework was proposed in our recent study \cite{feng2020unsupervised}, 
and showed state-of-the-art performances on the subword discriminability task on two databases in UAM: ZeroSpeech 2017 \cite{dunbar2017zero} and Libri-light \cite{kahn2019librilight}. 

In this work, we expand and extend the work in \cite{feng2020unsupervised}. Specifically, we (1) compare the proposed approach to a supervised topline system that is trained on transcribed data of the target language; (2) compare the proposed approach with another cross-lingual knowledge transfer method  \cite{shibata2017composite};
(3) investigate the potential of our approach in relation to the amount of unlabeled training material by varying the data between $600$ hours  (as used in \cite{feng2020unsupervised}) and $6,000$ hours, and compare the models' performance to the topline model. Throughout our experiments, English is chosen as the target low-resource language. Its phoneme inventory and transcriptions are assumed unavailable during system development. Dutch and Mandarin are chosen as the two OOD languages for which phoneme inventories and transcriptions are available.

Unsupervised subword modeling is typically evaluated using overall performance measures, such as ABX  \cite{versteegh2015zero,dunbar2017zero}, purity \cite{I3EWang}, normalized mutual information (NMI) \cite{ondel2018bayesian}. These metrics, however, do not provide insights on the approaches’ ability of modeling individual phonemes or phoneme categories. As the ultimate goal beyond unsupervised subword modeling is to discover basic speech units that have a good consistency with the true phonemes of the target language, we, to the best of our knowledge for the first time in the literature, additionally present detailed analyses that explore the question of the effectiveness of the proposed approach to capturing phoneme and articulatory feature (AF) information of the target language. 
The analyses are based on the standard ABX error rate evaluation \cite{versteegh2015zero}, which we adapted for this work (see Section \ref{sec:method_analyze}), and consist of two parts, i.e., an analysis at the phoneme level and at the AF level. The analyses are aimed at investigating what phoneme and AF information is (not) captured by the learned subword-discriminative feature representation, which can be used to guide future research to improve unsupervised subword modeling as well as AUD. Moreover, we correlate the phoneme-level ABX error rates and the quality of the cross-lingual phone labels which are used to train our back-end DNN-BNF model in order to study why the proposed approach performs differently in capturing different target phonemes' information, and how the performance is affected by the quality of cross-lingual phone labels.

The remainder of this paper is organized as follows. Section \ref{sec:related_works} provides a review of related works on the unsupervised subword modeling task. In Section \ref{sec:usm_approach}, we provide a detailed description of the proposed approach to unsupervised subword modeling, and introduce comparative approaches to compare against our approach. Section \ref{sec:method_analyze} describes the methodology  used for the phoneme-level and AF-level analyses. Section  \ref{sec:exp_setup} introduces the experimental design of this study, while Section \ref{sec:exp_results} reports the results. Section \ref{sec:analysis_on_results} describes the setup for conducting the phoneme- and AF-level analyses, and discusses the results of the analyses. Finally, Section \ref{sec:conclusion} draws the conclusions.

\section{Related works}
\label{sec:related_works}
\subsection{Unsupervised learning techniques}
Clustering algorithms and self-supervised learning techniques are widely applied in zero-resource speech modeling. 
For instance, the clustering approach using the Dirichlet process Gaussian mixture model (DPGMM) \cite{chang2013parallel} has shown to outperform all other competitors \cite{chen2015parallel} in the Zero Resource Speech Challenge (ZeroSpeech) 2015 \cite{versteegh2015zero}. 
Follow-up studies focused on improving speaker invariance of the input features to DPGMM  \cite{heck2017feature} (best-performing in ZeroSpeech 2017 \cite{dunbar2017zero}), \cite{feng2018exploiting,Higuchi2019}, exploring multilingual DPGMM \cite{chen2017multilingual,yuan2017extracting}, and alleviating the “over-fragment” nature, i.e., DPGMM’s shortcoming in producing an excessive number of fine-grained clusters  \cite{wu2018optimizing,feng2019_TASLP}. 
$K$-means and HMMs were also investigated for frame clustering   \cite{ansari2017unsupervised,manenti2017unsupervised}.

In self-supervised learning for zero-resource speech modeling \cite{oord2017neural,hsu2017nips,oord2018cpc,Pascual2019,Chung2019,badino2014auto}, targets that a model is trained to predict are computed from the data itself \cite{doersch2017multi}. A typical self-supervised representation learning model is the vector-quantized variational autoencoder (VQ-VAE) \cite{oord2017neural}, which 
achieved a fairly good performance in ZeroSpeech 2017 \cite{chorowski2019unsupervised} and 2019 \cite{Tjandra2019}, and has become more widely adopted  \cite{Niekerk2020,Tobing2020Cyclic,Tjandra2020transformer} in the latest ZeroSpeech 2020 challenge \cite{Dunbar2020zero}.
Other self-supervised learning algorithms such as factorized hierarchical VAE  (FHVAE) \cite{Hsu2018extracting}, contrastive predictive coding (CPC) \cite{oord2018cpc} and APC \cite{Chung2019} were also extensively investigated in unsupervised subword modeling \cite{Feng2019improving,riviere2020unsupervised,feng2020unsupervised,Niekerk2020} \textcolor{black}{ as well as in a relevant zero-resource word discrimination task \cite{vanStaden2021comparison}}.


\subsection{Cross-lingual knowledge transfer}
Transcribed speech data for OOD languages \cite{panayotov2015librispeech,aidatatang}
can be exploited in various ways to boost zero-resource subword modeling for low-resource languages. In \cite{shibata2017composite,feng2018exploiting}, a DNN AM trained with  OOD languages was used to extract cross-lingual phone posteriorgrams \cite{shibata2017composite} or cross-lingual BNFs \cite{shibata2017composite,feng2018exploiting} of a target low-resource language as the learned feature representation. 
In \cite{feng2018exploiting,feng2019_TASLP}, an OOD ASR system was used to generate phone labels for the target language speech. These cross-lingual labels served as supervision for training a DNN-BNF model. This idea is applied in our present study.

There is evidence that the above-mentioned cross-lingual phone labels are complementary to labels obtained with unsupervised learning    \cite{feng2018exploiting,feng2019_TASLP}.  Specifically, cross-lingual phone labels and DPGMM clustering labels for the target language’s speech data were jointly used to train a DNN-BNF.
The resulting BNF representation performed better than that extracted by a DNN-BNF trained using either type of the labels. 
\textcolor{black}{The work in \cite{hermann2021multilingual} adopted another way of combining unsupervised and cross-lingual learning strategies; they used cross-lingual BNFs as input to train a correspondence autoencoder (cAE).}
In our present study, the combination of unsupervised learning techniques and cross-lingual knowledge transfer is done in a different way by adopting a self-supervised learning front-end followed by a cross-lingual phone-aware DNN-BNF back-end. 


\subsection{Analysis of unsupervisedly discovered speech units}
Few analyses on the effectiveness of subword modeling at the phoneme level or of the linguistic relevance of the speech units learned using AUD exist \cite{feng2017linguistic,harwath2019towards,matusevych2020evaluating}. 
In \cite{feng2017linguistic}, an analysis on the consistency between individual discovered speech units from an unknown language and the language’s true phoneme inventories showed that while the learned speech units had a good coverage of the phoneme inventories, some phonemes with rapidly changing acoustics (e.g., diphthongs) could not be well discovered. In contrast to  \cite{feng2017linguistic},  our analysis study is based on the subword-discriminative feature representation instead of based on a discrete set of learned units. Moreover, our study also carries out performance analysis at the AF level in addition to the phoneme level. 
\textcolor{black}{A recent study \cite{matusevych2020evaluating} carried out an  analysis of the subword-discriminative feature representations, similar to our present study,  but with the purpose of comparing the unsupervised learning approaches to infant phonetic perception. A major difference between \cite{matusevych2020evaluating} and ours is that \cite{matusevych2020evaluating} selected only three phone contrasts from a target language, whereas our study conducts analysis on the complete phoneme inventories.
Finally, an analysis by \cite{harwath2019towards} showed that  a visually-grounded model (not requiring transcribed data) learns  subword units that carry AF information, such as  vowel backness or stop voicing. }



 




\section{Proposed approach to unsupervised subword modeling}
\label{sec:usm_approach}

\begin{figure}[!t]
    \centering
    \includegraphics[width=\linewidth]{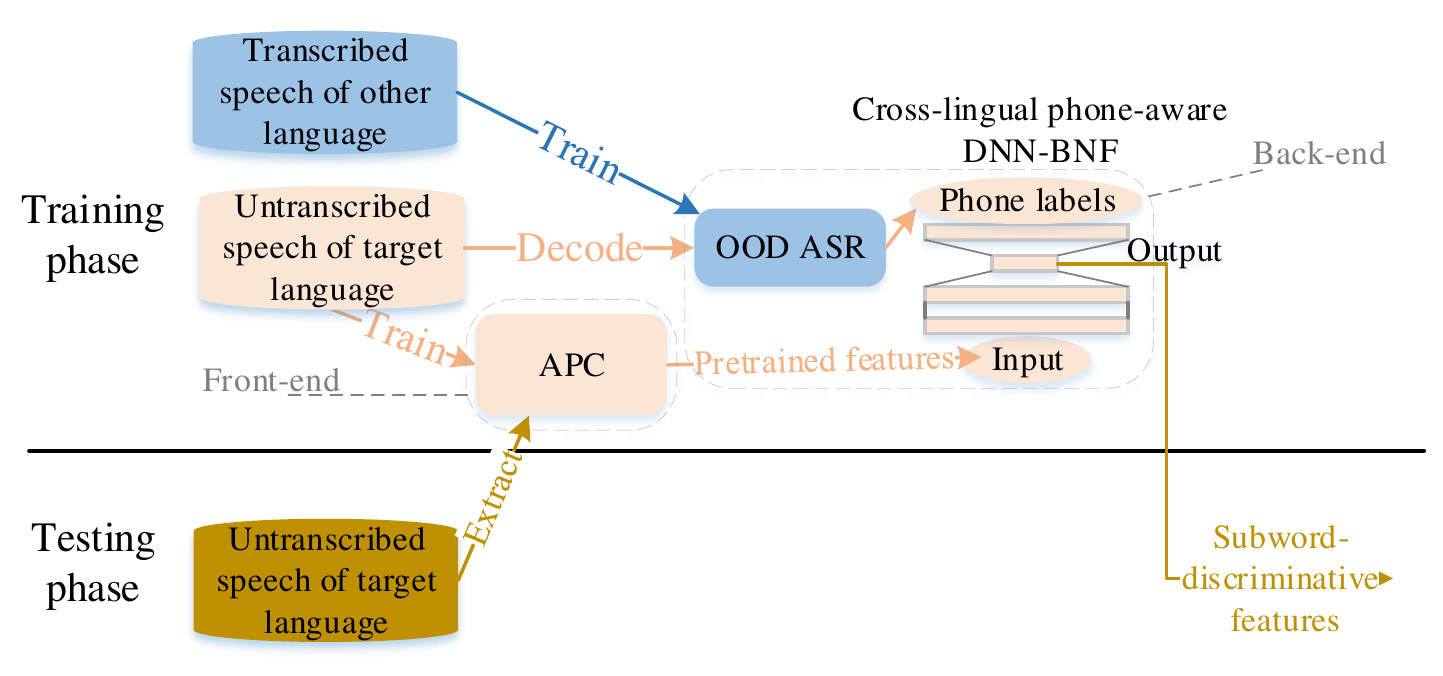}
    \caption{General framework of the proposed approach. The two colors in the training phase represent the data sets used to train each model.}
    \label{fig:general_framework}
\end{figure}

The general framework of the proposed approach to unsupervised subword modeling is illustrated in Fig. \ref{fig:general_framework}. In the training phase, for a target language with a certain amount of untranscribed speech training data (marked in pink in Fig. \ref{fig:general_framework}), 
an APC model as the front-end is trained with target  untranscribed speech in a self-supervised manner, and used to extract pretrained features for the target speech. 
Next, at the back-end, an OOD ASR system assigns one phone label to every frame of the target language’s speech. This OOD ASR is trained on a language different from the target language (marked in blue). With APC pretrained features for the target untranscribed speech as input features and cross-lingual phone labels as their corresponding target labels, a DNN-BNF model is trained, in order to learn subword-discriminative features for the target speech. In the testing phase, the DNN-BNF model is used to extract BNF representations for test data of the target language (marked in yellow) as the subword-discriminative feature representation of the target language’s speech.  

This study compares the proposed approach with other, state-of-the-art approaches \textcolor{black}{(see Section III-C for details)}. The front-end APC pretraining method is compared with another self-supervised learning method, i.e., FHVAE \cite{hsu2017nips}. Their comparison is made at  front-end level.
The whole pipeline of the proposed approach is compared with a system consisting of only the back-end DNN-BNF model (without an APC front-end), a CPC approach \cite{oord2018cpc}  and a transfer learning BNF system based on a cross-lingual AM  \cite{shibata2017composite}. Moreover, two different languages will be used to train two different OOD ASR systems for comparison.

\subsection{APC pretraining}
\label{subsec:approach_apc}
In previous studies, feature representation learning techniques were often adopted in order to suppress speaker variation while retaining linguistic information \cite{heck2017feature,feng2018exploiting,feng2019_TASLP}. 
\textcolor{black}{In contrast, APC adopted in the proposed approach is aimed at learning a frame-level feature representation that retains both phonetic and speaker information from the speech signal, while making the phonetic information and  speaker information more separable for downstream  phone or speaker classification tasks, comparing to when spectral features are used as frame-level representations \cite{Chung2019}.}
In such a way, the learned representation is considered to be less at risk of losing phonetic information compared to that learned by methods in \cite{heck2017feature,feng2018exploiting,feng2019_TASLP}.

Let $\{\bm{x_1},\bm{x_2}, \ldots, \bm{x_T} \}$ denote $d$-dimensional frame-wise features for a set of untranscribed speech data for APC training, where $T$ is the total number of speech frames. 
At each time step $t$, the encoder of APC, denoted as $\mathrm{Enc} (\cdot)$, reads as input a feature vector $\bm{x_t}$, and generates a $d$-dimensional output feature vector $\bm{\hat{x}_t}$ based on 
previous input features $\bm{x_{1:t}} = \{\bm{x_1}, \bm{x_2}, \ldots, \bm{x_t}\}$, 
\begin{equation}
    \bm{\hat{x}_t} = \mathrm{Enc} (\bm{x_{1:t}}).\label{eqt:enc}
\end{equation}
The goal of APC is to let $\bm{\hat{x}_t}$ be as close to $\bm{x_{t+n}}$ as possible, where $n$ is a pre-defined constant non-negative integer, known as the \textit{prediction step}. The loss function for APC training is defined as,
\begin{equation}
    \mathrm{Loss} = \sum_{t=1}^{T-n} \left| \bm{\hat{x}_t}  - \bm{x_{t+n}}\right|.
\end{equation}
With this loss function, the APC encoder learns information that is relevant to predicting future frames. 
Intuitively, a large $n$ encourages the APC model to encode  global characteristics in the original speech signal, while a small $n$ lets the model   focus mainly on local smoothness in speech. 

The APC encoder is implemented as a long short-term memory (LSTM) RNN structure \cite{sak2014long}, as was done in \cite{Chung2019}. 
Let $L$ denote the number of LSTM layers,  Equation (\ref{eqt:enc}) is formulated as,
\begin{align}
    \bm{h_0} &= \bm{x_{1:t}},    \\
    \bm{h_l} &= \textrm{LSTM}^l (\bm{h_{l-1}}), l\in \{1,2,\ldots, L\}, \\
    \bm{\hat{x}_t} &= \bm{W} \bm{h_L},
\end{align}
where $\bm{W}$ is a trainable projection matrix. The equations that form  $\textrm{LSTM} (\cdot)$ can be found in \cite{sak2014long}. 



After APC training, $\bm{h^{L}} = \{\bm{h^{L}_{1:T}}\}$ is extracted as the learned acoustic representation of the original speech, and is henceforth referred to as the \textit{APC feature}. In principle, $\bm{h^l}$ of an arbitrary layer could be extracted as the learned representation. We follow \cite{Chung2019} in using the representation from the top layer as they showed that this gave the optimal performance on phone classification tasks.

\subsection{Cross-lingual phone-aware DNN-BNF}
\label{subsec:approach_crs_ling_bnf}
As shown in Fig. \ref{fig:general_framework}, the back-end of the proposed approach is a DNN model with a low-dimensional intermediate hidden layer, also known as the bottleneck layer. To train such a DNN-BNF model, cross-lingual phone labels are obtained beforehand. Specifically, an OOD ASR system is applied to decode speech utterances of the target language’s training data. 
The decoding results, i.e. hypothesized transcripts are generated for speech utterances. 
By applying forced alignment to the hypothesized transcripts \textcolor{black}{(i.e. OOD phone sequences as labels)} using the AM of the OOD ASR system, each frame of the training utterance is assigned a phone symbol label modeled by the OOD ASR.
In this work the OOD ASR is realized as a hybrid DNN-HMM architecture. 
As a result, the cross-lingual phone labels are triphone HMM states\footnote{\textcolor{black}{In the literature, they are also   referred to as \textit{senones}}.} modeled by the hybrid DNN-HMM. In principle, ASR systems with other architectures could also be applied to generate decoding-based labels for target unlabeled speech, such as connectionist temporal classification (CTC)  \cite{graves2014towards} and attention-based models \cite{chan2016listen}.

The DNN-BNF model is then trained with the speech data of the target language, using the pretrained APC features as input features and the cross-lingual labels as target labels. The training is done by minimizing cross-lingual phone prediction error by using the lattice-free maximum mutual information (LF-MMI) criterion \cite{povey2016purely}. 
After training, the DNN-BNF model is used to extract the BNF representations of the test data as the desired subword-discriminative feature representation. The BNF representation is essentially the output of the bottleneck layer, as shown in Fig. \ref{fig:general_framework}.

\subsection{Comparative approaches}
\label{subsec:approach_comparative_approaches}
\subsubsection{FHVAE}
 FHVAE  is a self-supervised representation learning model which does not need  transcribed  data in model training \cite{hsu2017nips}. It
disentangles phonetic and speaker information by capturing the two types of information with latent sequence variables $\bm{z_1}$ and latent segment variables $\bm{z_2}$ respectively. The  $\bm{z_1}$ representation from a well-trained FHVAE is extracted as the desired speaker-invariant  phonetic representation for unsupervised subword modeling.
The FHVAE model was applied in    \cite{Feng2019combining} and achieved good performance in the ZeroSpeech 2019 Challenge \cite{Dunbar2019}, which is why we compare the APC model against FHVAE in this study.
Details of the FHVAE model description is provided in supplementary material (see Section \ref{subsec:supple_fhvae_dtails}).

\subsubsection{Cross-lingual AM based BNF}
\label{subsubsec:approach_comparative_approaches_crsling_am}
Learning subword-discriminative feature representations by exploiting   cross-lingual knowledge transfer could be realized in a different way than the back-end of our proposed approach discussed in Section  \ref{subsec:approach_crs_ling_bnf}. In our back-end, the DNN-BNF model trained using unlabeled speech data of the target language and cross-lingual phone labels is used to extract the BNF representation. Alternatively, a DNN-BNF model trained using labeled speech data of an OOD language can be leveraged to extract the BNF representation for the target language  \cite{shibata2017composite,tsuchiya2018speaker}. 
In essence, the method in \cite{shibata2017composite,tsuchiya2018speaker}
leverages a well-trained cross-lingual AM for transfer learning. Thus this method is denoted as the \textit{cross-lingual AM based BNF}, to be distinguished from  our back-end  cross-lingual phone labeling based method. 
The cross-lingual AM based BNF method does not rely on  audio data of the target language for training, which makes it fast in system development. Moreover, this method is feasible in a stricter zero-resource scenario where unlabeled audio data of the target language is unavailable for training. We will compare our entire approach pipeline against the cross-lingual AM based BNF method. 

\subsubsection{CPC}
CPC  is a  self-supervised  representation  learning  model which does not require transcribed data in model training \cite{oord2018cpc}. By  using  a contrastive  loss,  the  model  is trained  to  distinguish  future speech  frames  from  a  set  of  negative  examples.  The CPC model is able to capture  phonetic information  in  speech  while  suppressing  noise  and  speaker variation \cite{kahn2019librilight}, and achieves good performance  in  unsupervised  subword modeling \cite{kahn2019librilight,riviere2020unsupervised,Kharitonov2020data_augment}. This is why we compare our approach to CPC.



\section{Methods to analyze the effectiveness of the proposed approach}
\label{sec:method_analyze}
This section describes our phoneme-level and articulatory feature-level methods to analyze the effectiveness of the proposed approach to unsupervised subword modeling. Both methods are based on the ABX test   \cite{munson1950standardizing}.   Section  \ref{subsec:analysis_abx_overall} briefly introduces the ABX test and its application as an overall performance measure in unsupervised subword modeling. Sections \ref{subsec:analysis_phoneme_level} and \ref{subsec:analysis_af_level} discuss the proposed phoneme- and AF-level analyses methods respectively.
\subsection{ABX   subword discriminability task}
\label{subsec:analysis_abx_overall}
The ABX test was first proposed in   \cite{munson1950standardizing} as as a human auditory test. $A, B$ and  $X$ are three audio signals. A listener is presented with $A, B$ and $X$, and is asked to make judgement about whether $X$ is more similar to $A$ or to $B$. 

The ABX test has recently been adopted as one of the standard evaluation metrics in unsupervised subword modeling  \cite{versteegh2015zero,dunbar2017zero,Dunbar2019}.
Specifically, $A, B$ and $X$ are feature representations of three speech sounds.  $A$ and $B$ contain triphone sequences that differ only in the central phone, e.g. ``a-p-i'' versus ``a-t-i''. $X$ contains either ``a-p-i'' or ``a-t-i''. 
The ABX error rate of a pair of triphone sequences $x$ and $y$ is defined as,
\begin{equation}
    \epsilon (x, y) = \frac{1}{2}[ \textcolor{black}{\eta(x \rightarrow y)} + \textcolor{black}{\eta (y \rightarrow x)}],
    \label{eqt:abx_error_rate_triphones}
\end{equation}
where 
\begin{equation}
\begin{split}
 \textcolor{black}{\eta(x \rightarrow y)} =& \frac{1}{\vert S(x) \vert (\vert S(x) \vert-1) \vert S(y) \vert} \sum_{A \in S(x)} \sum_{B \in S(y)}\sum_{X \in S(x) \backslash \{A\}} \\
 &(\mathbb{1}_{d(A,X) > d(B,X)} + \frac{1}{2} \mathbb{1}_{d(A,X) = d(B,X)}).
\end{split}
\label{eqt:abx_error_rate_triphone_asym}
\end{equation}
Here $\mathbb{1}$ is the indicator function, \textcolor{black}{$S(x)$ and $S(y)$ denote two sets of speech sounds containing $x$ and $y$}, $d(\cdot, \cdot)$ denotes dynamic time warping (DTW) based dissimilarity between two speech sounds. Frame-level dissimilarity measure for DTW scoring can be the cosine distance, Kullback–Leibler (KL) divergence, etc. 
\textcolor{black}{Here, cosine distance is used throughout this paper.}
By taking average of $\epsilon(x,y)$ over all possible triphone sequences $x$ and $y$ that share context phones and contrast the same central phone pair, the phone \textbf{pairwise ABX error rate} is calculated. By further taking the average of the phone pairwise ABX error rates over all possible pairs of phones, the \textbf{overall ABX error rate} is calculated.



\subsection{Phoneme-level analysis}
\label{subsec:analysis_phoneme_level}
We define   \textbf{phoneme-level  ABX error rate}  as follows. 
Let $\omega$ be a phoneme in phoneme inventory $\Omega$ of a target language. The phoneme-level ABX error rate of $\omega$ is then calculated as,
\begin{equation}
 \xi(\omega) =  \frac{1}{\left| \Omega \right| -1}  \sum_{\omega^{\prime} \in \Omega \backslash \{\omega\} } \epsilon(\omega, \omega^{\prime}),
\end{equation}
where  $\epsilon(\omega, \omega^{\prime})$ is the pairwise ABX error rate calculated as mentioned in \ref{subsec:analysis_abx_overall}. 
$\xi(\cdot)$ is used as the measure of a subword-discriminative feature representation’s effectiveness towards each individual phoneme in a target language. 

The present study also investigates the correlation between phoneme-level ABX error rate improvement achieved by the back-end of the proposed approach and the quality of the cross-lingual phone labels used in the back-end.
Let $\Psi = \{\psi_1, \ldots, \psi_M\}$ denote $M$ cross-lingual phones modeled by an OOD ASR system, and $\{l_t \in \Psi | t=1,2,\ldots, T\}$ denote cross-lingual phone labels for a certain amount of target speech data generated by the OOD ASR.  Let $\{g_t \in \Omega | t=1,2,\ldots, T \}$ denote the true phoneme labels for the same speech data, where $\Omega = \{\omega_1, \ldots, \omega_N\}$ is the set of $N$ phonemes in the target language.  An $N$-by-$M$ confusion matrix $\bm{E}$ can be constructed, with its element $e_{ij}$ defined as,
\begin{equation}
    e_{ij} = \frac{\sum_{t=1}^{T} \mathbb{1} (g_t = \omega_i, l_t =\psi_j)}{\sum_{t=1}^{T} \mathbb{1} (g_t = \omega_i)}.
    \label{eqt:confusion_matrix}
\end{equation}
Conceptually, $e_{ij} \in [0,1]$ represents the percentage of target speech frames of true phoneme $\omega_i$ that are labeled as cross-lingual phone $\psi_j$. 
As seen from Equation (\ref{eqt:confusion_matrix}), elements in every row of $\bm{E}$ sum up to $1$. Ideally the row vector $[e_{i,1},e_{i,2},\ldots, e_{i,M}]$ is a (quasi) one-hot vector, in which case there exists a cross-lingual phone $\psi_{j*}$ so that $e_{ij*}$ is close to $1$. 
A large $e_{ij*}$ indicates 
  high consistency between  $\omega_i$ and $\psi_{j*}$, and is desired in the cross-lingual DNN-BNF model training, as the DNN-BNF model can be trained to map acoustic features representing different speech realizations of $\omega_i$ to very close representations in the cross-lingual phonetic space.  It is worth noting that the consistency being discussed here measures to which extent speech frames of $\omega_i$ get the same cross-lingual phone labels irrespective of the labels’ symbol.

Mathematically,   $j* = \argmax_{j=1}^{M} e_{ij}$, and the co-occurrence probability between $\omega_i$ and $\psi_{j*}$, denoted as $p_{co} (\omega_i)$, is defined as,
\begin{equation}
     p_{co} (\omega_i) = \max_{j=1}^{M} e_{ij}.
    \label{eqt:p_co}
\end{equation}
In this study, $p_{co} (\omega_i)$ is  utilized  as a measure of the  cross-lingual phone label quality of $\omega_i$.




\subsection{AF-level analysis}
\label{subsec:analysis_af_level}
An AF-level analysis is carried out using the proposed \textit{ABX  AF discriminability task} in order to evaluate how well a speech feature representation is capable of distinguishing one AF attribute from another.
Analogous to the ABX subword discriminability task as introduced in Section \ref{subsec:analysis_abx_overall}, let $A$ and $B$ contain triphone sequences that differ only in the central phone. Here the central phones of  $A$ and $B$ are set to belong to different \textbf{AF attributes}. For instance, take manner of articulation (MoA): $A$ could contain a  \textit{stop} in the center (e.g., A could be /a p i/), 
while  $B$  contains a \textit{fricative} (e.g., /a f i/). $X$ contains a triphone sequence with its central phone being either a  stop or a fricative, but  not necessarily  ``p'' or ``f''.
The context phones of $X$ are ``/a/'' and ``/i/''. Possible realizations of $X$ could be ``a g i'', ``a z i'', etc.
Using Equations (\ref{eqt:abx_error_rate_triphones}) and (\ref{eqt:abx_error_rate_triphone_asym}), followed by taking the average over all possible triphone sequences that share the context phones and contrast the stop and fricative attributes in the central phones, the \textbf{pairwise ABX AF error rate} between stops and fricatives is calculated.
The \textbf{attribute-level ABX AF error rate} (e.g. the stop attribute) can then be calculated by taking the average of all pairwise ABX AF error rates involving that AF attribute.


The present study carries out AF-level analysis on two AFs for consonants, i.e.,  MoA and place of articulation (PoA), and on two AFs for vowels, i.e., tongue height and tongue backness. The mappings from an English phoneme to its AF attributes are shown in Tables \ref{table:AF_moa_poa} (consonants) and \ref{table:AF_backness_height} (vowels). 
In these two tables, each phoneme is represented as its ARPABET symbol \cite{wiki:arpabet}.
For the  analyses of tongue height and backness, all diphthongs are excluded, 
because it does not have a stable tongue height or backness attribute but rather during its production the articulators move from one position to another.



\begin{table}[!t]
\renewcommand\arraystretch{0.5}
\centering
\caption{ MoA (rows) and PoA (columns) for each English consonant.
 PoA abbreviations from left to right: Bilabial, Labiodental, Dental,  Alveolar, Postalveolar, Palatal, Velar, Glottal. MoA abbreviations from top to bottom: Affricate, Approximant, Fricative, Stop, Nasal.}
\resizebox{   \linewidth}{!}{%

\begin{tabular}{l|cccccccc}
\toprule
\textbf{Consonants} &  BI  &  LA  &  DA  &  AL  &  PO  &  PA  &  VE  &  GL  \\ 
\midrule
Af  &           &            &           & &  CH, JH     &         &           &           \\ 
\midrule
Ap  & W         &            &           & L & R       & Y         &       &           \\
\midrule
Fr   &           & F, V       & TH, DH     & S, Z & SH, ZH   &      &           & HH        \\ 
\midrule
St   & P, B &            &           & T, D &         &           & K, G       &           \\ 
\midrule
Na  & M         &            &           & N &         &           & NG        &           \\ 

\bottomrule
\end{tabular}
}
\label{table:AF_moa_poa}
\end{table}

\begin{table}[!t]
\centering
\renewcommand\arraystretch{0.5}
\caption{Tongue height (rows) and backness (columns) for each English monophthong vowel. Diphthongs are excluded in this study.}
\begin{tabular}{l|ccc}
\toprule
\textbf{Vowels} & Front & Central & Back\\ 
\midrule
Close    & IY, IH & &  UW, UH   \\
Mid &EH &ER, AH &AO\\
Open &AE &AA &\\
\bottomrule
\end{tabular}
\label{table:AF_backness_height}
\end{table}

\section{Experimental setup}
\label{sec:exp_setup}
\subsection{Databases and evaluation metric}
Training data for APC and DNN-BNF models are taken from Libri-light \cite{kahn2019librilight}, a recently developed and freely-available English database to support unsupervised subword modeling. The \textit{unlab-600} and \textit{unlab-6K} sets from Libri-light are adopted. Details of the two training sets are listed in Table   \ref{tab:exp_setup_libri_light_no_subsets}. 



\begin{table}[!t]
\renewcommand\arraystretch{0.5}
\centering
\caption{Details of two  training sets in Libri-light.}
\begin{tabular}{l|ccc}      
\toprule
& \#utterances & \#speakers & \#hours \\
\midrule
unlab-600 & $35,229$ & $489$ & $526$ \\
unlab-6K  & $362,816$  & $1,742$ & $5,273$ \\

\bottomrule
\end{tabular}%
\label{tab:exp_setup_libri_light_no_subsets}
\end{table}

Dutch and Mandarin training data for the development of two OOD ASR systems are the Corpus Gesproken Nederlands (CGN) \cite{oostdijk2000spoken} and Aidatatang\_200zh (ADT) \cite{aidatatang} respectively. The training-test partition of CGN follows those in \cite{laurensw75cgn_kaldi}. Its training data contains $483$ hours of speech covering conversational and read speech and broadcast news. ADT is a read speech corpus. Its training data consists of $140$ hours of speech.

Evaluation data are taken from Libri-light and ZeroSpeech 2017 Track 1 \cite{dunbar2017zero}. The Libri-light evaluation data consists of $4$ sets: \textit{dev-clean, dev-other, test-clean, test-other}. Speech in dev-clean and test-clean have higher recording quality and accents closer to US English than that in dev-other and test-other.

The ZeroSpeech 2017 evaluation data consists of three languages, i.e., English, French and Mandarin \cite{dunbar2017zero}. In this study, the English evaluation data is chosen. The data are organized into subsets of differing lengths (1s, 10s and 120s). 

The learned BNF representations, as well as APC pretrained feature representations are evaluated in terms of the overall ABX error rate (see Section \ref{subsec:analysis_abx_overall}), for \textit{within-speaker} ($X$ and $A/B$ are spoken by the same speaker) and \textit{across-speaker} ($X$ and $A/B$ are spoken by different speakers) separately. The across-speaker ABX task is more challenging, and is particularly focused on measuring the robustness of learned feature representations towards speaker variation. \textcolor{black}{The within-speaker task serves as a benchmark for the across-speaker task by showing to what extent the ABX error in the across-speaker performance is caused by speaker variation.}


\subsection{Baselines and toplines}
\label{subsec:exp_setup_baseline_topline}
The official baseline systems of Libri-light and ZeroSpeech 2017   \cite{kahn2019librilight,dunbar2017zero} both use raw MFCC features.  
The official topline of ZeroSpeech 2017 is a supervised system which uses English labeled data to train an ASR system, followed by generating phone posteriorgram as the learned feature representation \cite{dunbar2017zero}.

There is no official topline system provided for the Libri-light database. In this work we created a supervised system and used it as the topline of Libri-light. This system uses 960 hours of English labeled data from Librispeech \cite{panayotov2015librispeech} to train a time-delay NN (TDNN) AM, from which the output of the top TDNN layer is used as the learned representation. 
Training of the TDNN AM is implemented based on the Kaldi Librispeech recipe\footnote{\texttt{s5/local/nnet3/run\_tdnn.sh}.} without modifying any parameters. The TDNN AM consists of five $650$-dimensional hidden layers.
The input to the TDNN consists of $40$-dimensional high-resolution MFCC features (HR MFCCs) appended by $100$-dimensional i-vectors. Frame labels needed to train the TDNN AM are obtained by forced-alignment with a GMM-HMM AM trained beforehand, also using Librispeech.


\subsection{Front-end implementation}
The front-end APC model is implemented as a multi-layer LSTM network. Residual connections are made between two consecutive layers. Each layer has $100$ dimensions. 
In order to find the optimal number  of LSTM layers and prediction step $n$, layer numbers ranging in $\{3,4,5,6\}$, with   $n$ ranging in $\{1,2,3,4,5\}$\footnote{Increasing $n$ to larger than $5$ was found leading to rapid ABX error rate degradation in preliminary experiments.} are  tested when the model is trained with unlab-600. These results are reported in supplementary material (see Section \ref{subsec:supple_apc_para}).
\textcolor{black}{From Table \ref{tab:supple_exp_results_apc_paras} the optimal parameters of LSTM layer number and $n$ can be determined as $5$ and $5$ respectively.}
The two parameters are then fixed and the APC  model is  trained with the unlab-6K set from scratch. The input features to APC are $13$-dimension MFCCs with cepstral mean normalization (CMN). The APC models are trained using an open-source tool by \cite{Chung2019} for a fixed $100$ epochs throughout our experiments, with the Adam optimizer \cite{kingma2014adam}, an initial learning rate of $0.0001$, and a batch size of $32$. After training, output of the top LSTM layer is extracted as the APC feature representation. 

\subsection{Back-end implementation}
\subsubsection{OOD ASR systems}
\label{subsubsec:expsetup_backend_ood}
Two OOD ASR systems were developed, i.e., a Dutch ASR and a Mandarin ASR. Both OOD ASR systems use a chain TDNN AM trained in Kaldi \cite{povey2011kaldi}  and a tri-gram LM trained using SRILM toolkit \cite{Stolcke02srilm--}. The TDNN AM contains $7$-layers, including a $40$-dimension bottleneck layer at the $6$-th layer. Three-way speed perturbation \cite{ko2015audio} is applied to the Dutch (CGN) and Mandarin (ADT) training data respectively, before they are used to train the TDNN AM.
The model is trained based on the lattice-free maximum mutual information (LF-MMI) criterion \cite{povey2016purely}. For Dutch, the input features are $40$-dimension HR MFCCs. For Mandarin, the input features consist of HR MFCCs appended by pitch features (MFCC+P) \cite{ghahremani2014pitch}. Frame labels for TDNN model training are obtained by forced alignment using a GMM-HMM AM trained beforehand. 
\textcolor{black}{The numbers of modeled 
phones and triphone HMM states in the Dutch TDNN AM are $81$ and $3,361$, respectively. The number of modeled phones in the Mandarin TDNN AM is $119$,  including tone-dependent phones that are used to describe the four tones plus a neutral tone in Mandarin. The number of   triphone HMM states in the Mandarin TDNN AM is $3,536$.}

The Dutch ASR obtained a word error rate (WER) of $8.98\%$ on the CGN broadcast test set. (This WER could be improved upon by integrating an RNN LM. As Dutch ASR performance is not the focus in this work, an RNN LM is not applied). The Mandarin ASR obtained a character error rate (CER) of $6.37\%$ on the ADT test set. The two ASR systems are used to generate cross-lingual phone labels for the speech frames in Libri-light unlab-600 and unlab-6K sets.
\subsubsection{DNN-BNF model}
\label{subsubsec:expsetup_backend_dnn_bnf}
Two DNN-BNF models are trained, one taking the Dutch phone labels as training labels and one taking the Mandarin phone labels as training labels.

The DNN-BNF model consists of $7$ feed-forward layers. Each layer has $450$ dimensions except a $40$-dimensional bottleneck layer, which is located below the top layer. The DNN-BNF model is trained based on the LF-MMI criterion. The inputs to the DNN-BNF are the APC feature with its neighboring frames, ranging from $-3$ to $+3$. After training, $40$-dimensional BNFs are extracted, and are evaluated in terms of overall ABX error rate.. The BNF representations are named as \textbf{A-BNF-Du} and \textbf{A-BNF-Ma} henceforth, where ``-Du'' and ``-Ma'' denote using the Dutch and Mandarin training labels respectively, and “A-” denotes APC features as input features. 

For comparison, two other DNN-BNF models (one using the Dutch phone labels as training labels and one using the Mandarin phone labels) are trained using HR MFCCs as input features.
Other training and model parameter settings are unchanged. After training, again the BNFs are extracted and evaluated. The BNF representations are henceforth named as\textbf{M-BNF-Du} and \textbf{M-BNF-Ma}, where “M-” stands for taking MFCC features as input features.

\subsection{Implementation of the comparative approaches}
\subsubsection{FHVAE}
Model parameters of FHVAE are chosen by referring  to \cite{Feng2019improving}. The FHVAE encoder and decoder are implemented as $2$-layer LSTM networks. Each LSTM layer has $256$ dimensions. Latent segment variable $\bm{z_1}$ and latent sequence variable $\bm{z_2}$ are both $32$ dimensional. The inputs to FHVAE are fixed-length speech segments, each of which consists of $20$ frames. Frame-level input features are MFCC with CMN. The FHVAE models are trained using an open-source tool by \cite{hsu2017nips}, with the Adam optimizer \cite{kingma2014adam}. The model is trained with unlab-600 in Libri-light. A $10\%$ subset of the training data is randomly selected for cross-validation (CV). The training procedure terminates if FHVAE’s lower bound on the CV set does not improve for $40$ epochs. After training, $\bm{z_1}$ is extracted.
\subsubsection{Cross-lingual AM based BNF}
The cross-lingual AM based BNF representation is generated based on the TDNN AM of the OOD ASR systems described in Section \ref{subsubsec:expsetup_backend_ood}.   To that end, speech features of the English evaluation data are fed to the OOD (Dutch or Mandarin) TDNN AM till its bottleneck layer. In this way, two BNF representations are extracted, one from the Dutch TDNN AM, and one from the Mandarin TDNN AM.  The two BNF representations are named \textbf{C-BNF-Du} and \textbf{C-BNF-Ma}, respectively, where ``C-'' 
stands for the  comparative approach. 

To give an explicit comparison between the cross-lingual AM based BNFs and  BNFs from our proposed approach (in Section \ref{subsubsec:expsetup_backend_dnn_bnf}), Table \ref{table:expsetup_naming} lists their configurations.
The row ``Acoustic'' denotes the acoustic training data, the rows ``Input'' and ``Label'' denote input feature representation and frame labels used to train the respective systems. 


\begin{table}[!t]
\centering
\renewcommand\arraystretch{0.50}
\caption{  Configurations of the BNF representations implemented in the experiments.}
\resizebox{  \linewidth}{!}{%
\begin{tabular}{l|c|c|c|c|c|c}
\toprule
 & M-BNF-Du  & M-BNF-Ma  &  A-BNF-Du  &  A-BNF-Ma  & C-BNF-Du  &    C-BNF-Ma    \\ 
\midrule

Method & \multicolumn{4}{c|}{cross-lingual phone labeling} & \multicolumn{2}{c}{cross-lingual AM}\\
\midrule
Acoustic& \multicolumn{4}{c|}{Libri-light} & CGN & ADT  \\
\midrule
Input&\multicolumn{2}{c|}{MFCC} & \multicolumn{2}{c|}{APC} & MFCC & MFCC+P\\
\midrule
Label & Du. ASR & Ma. ASR & Du. ASR & Ma. ASR & CGN  & ADT \\
\bottomrule
\end{tabular}
}
\label{table:expsetup_naming}
\end{table}
\section{Experimental results}
\label{sec:exp_results}
\subsection{Effectiveness of the front-end APC pretraining}
\label{subsec:exp_results_front_end}

First,  the APC and FHVAE methods are compared at the front-end, i.e., without using  the DNN-BNF back-end. Overall across-speaker and within-speaker ABX error rates  ($\%$) of the APC features,  the FHVAE features, and the official MFCC baseline  \cite{kahn2019librilight} on the four Libri-light evaluation sets are listed in Table  \ref{tab:exp_results_apc_paras} separately and averaged over all evaluation sets (right-most column). 
The APC models in this table are trained with unlab-600\footnote{APC trained with unlab-600 was also published in our recent study \cite{feng2020unsupervised}.} and unlab-6K respectively, and FHVAE is trained with unlab-600.
\begin{table}[!t]
\renewcommand\arraystretch{0.60}
\centering
\caption{Overall ABX error rates ($\%$) of the APC features, FHVAE features, and the official MFCC baseline on Libri-light. Bold indicates the best result. Data within brackets indicates the training set for each APC and FHVAE model.}
\resizebox{1 \linewidth}{!}{%
\begin{tabular}{l|cccc|c}      
\toprule
\midrule[0.2pt]
\multicolumn{6}{c}{Across-speaker}\\
System & dev-clean & dev-other & test-clean & test-other & Avg. \\
\midrule
APC (unlab-600) \cite{feng2020unsupervised} & 


 $12.64$&$19.00$&$12.19$&$18.75$&$15.65$\\


APC (unlab-6K) &$10.79$&$16.87$&$10.33$&$16.79$&$\bm{13.70}$\\
\midrule
MFCC baseline \cite{kahn2019librilight}  & $20.94$&	$29.41$&	$20.45$&	$28.50$&	$24.83$\\
FHVAE (unlab-600)  &$18.41$&$25.66$&$17.79$&$25.65$&$21.88$\\

\midrule
 \multicolumn{6}{c}{Within-speaker} \\
\midrule
APC (unlab-600) \cite{feng2020unsupervised}  &


$8.83$&$11.07$&$8.36$&$11.48$&$9.94$\\
APC (unlab-6K)   & $7.49$&$9.99$&$7.05$&$10.11$&$\bm{8.66}$\\
\midrule
MFCC baseline \cite{kahn2019librilight}  & $10.95$&$13.55$&$10.58$&$13.60$&$12.17$\\
FHVAE (unlab-600) &$10.30$&$13.15$&$10.21$&$13.16$&$11.71$\\
\midrule[0.2pt]
\bottomrule
\end{tabular}%

}
\label{tab:exp_results_apc_paras}
\end{table}

The results show that when trained with unlab-600,  APC  outperforms both the FHVAE feature representations and the MFCC baseline. The lower across-speaker ABX error rate results for the APC features imply that they are more speaker invariant than the FHVAE features, even though the APC model is not explicitly suppressing speaker variation as FHVAE is.


Table \ref{tab:exp_results_apc_paras} shows for the APC model, when 
the amount of training data is increased ten-fold by training  on  unlab-6K, the APC feature representation further improves
by relative ABX error rate reduction of $12.5\%$ in the across-speaker condition and $12.9\%$ in the within-speaker condition.






\subsection{Effectiveness of the proposed approach}
\label{subsec:exp_results_back_end}

\begin{table}[!t]
\renewcommand\arraystretch{0.50}
\centering
\caption{Overall ABX error rates ($\%$) of the proposed cross-lingual phone-aware BNFs, comparative approaches, and supervised topline    on Libri-light. Systems listed under ``NOT in Libri-light'' used various different datasets other than unlab-600 or unlab-6K.}
\resizebox{  \linewidth}{!}{%
\begin{tabular}{l|cccc|c}      
\toprule
\midrule[0.2pt]
 \multicolumn{6}{c}{Across-speaker}\\
System& dev-clean & dev-other & test-clean & test-other & Avg.\\ 
\midrule
\multicolumn{6}{l}{\textit{Training set: unlab-600}}\\
M-BNF-Du  &$6.67$&$	11.65$&	$6.64$&$	12.00$&$	9.24$\\
M-BNF-Ma  & $7.92$&	$12.71$&	$7.74$&	$13.23$&	$10.40$\\
A-BNF-Du   & $\bm{6.18}$&	$\bm{11.02}$&$	\bm{6.03}$	&$\bm{10.94}$&$	\bm{8.54}$\\
A-BNF-Ma  & $7.00$&$11.80$&$6.84$&$11.81$&$9.36$\\


APC&   $12.64$&$	19.00$&$12.19$&$18.75$&$15.65$\\
CPC \cite{kahn2019librilight}  &$9.58$&	$14.67$&	$9.00$&	$15.10$&	$12.09$\\
\midrule
\multicolumn{6}{l}{\textit{Training set: \underline{unlab-6K}}} \\

M-BNF-Du & $6.06$&$11.30$&$6.12$&$11.35$&$8.71$ \\
M-BNF-Ma & $7.80$&$12.39$&$7.60$&$12.95$&$10.19$ \\
A-BNF-Du & $\bm{5.70}$&$\bm{10.08}$&$\bm{5.70}$&$\bm{10.02}$&$\underline{\bm{7.88}}$   \\
A-BNF-Ma & $6.38$&$11.02$&$6.23$&$11.02$&$8.66$ \\

APC&  $10.79$&$16.87$&$10.33$&$16.79$ &$13.70$\\
CPC \cite{kahn2019librilight} & $8.48$&$13.39$&$8.05$&$13.81$&$10.93$ \\
\midrule
\multicolumn{6}{l}{\textit{Training set \textbf{NOT} in Libri-light}}\\
Topline & $5.30$&$9.58$&$5.47$&$9.64$&$7.50$\\
CPC+DA \cite{Kharitonov2020data_augment} & $6.62$&$10.60$&$5.90$&$10.95$&$8.52$\\
C-BNF-Du & $7.17$&$11.20$&$6.89$&$11.40$&$9.17$\\
C-BNF-Ma & $9.92$&$14.91$&$9.83$&$15.34$&$12.50$\\
 \midrule
\multicolumn{6}{c}{Within-speaker} \\
\midrule
\multicolumn{6}{l}{\textit{Training set: unlab-600}}\\
M-BNF-Du&  $4.97	$&$6.94$&	$4.73$&	$6.86$&	$5.88$\\
M-BNF-Ma&  $6.06$&	$7.71$&	$5.62$&	$7.82$&	$6.80$\\
A-BNF-Du   & $\bm{4.77}$&	$\bm{6.69}$&	$\bm{4.49}$&	$\bm{6.43}$&	$\bm{5.60}$\\
A-BNF-Ma&  $5.25$&	$7.14$&	$5.21$&	$7.09$&	$6.17$\\
APC&  $8.83$&$11.07$&$8.36$&$11.48$&$9.94$\\
CPC \cite{kahn2019librilight}& $7.36$&	$9.39$&	$6.90$&	$9.59$&	$8.31$\\
\midrule
\multicolumn{6}{l}{\textit{Training set: \underline{unlab-6K}}}\\
M-BNF-Du & $4.70$&$6.58$&$4.36$&$6.37$&$5.50$\\
M-BNF-Ma & $5.94$&$7.65$&$5.69$&$7.77$&$6.76$\\
A-BNF-Du & $\bm{4.48}$&$\bm{6.15}$&$\bm{4.24}$&$\bm{5.91}$&$\underline{\bm{5.20}}$\\
A-BNF-Ma & $5.03$&$6.77$&$4.65$&$6.42$&$5.72$\\
APC & $7.49$&$9.99$&$7.05$&$10.11$&$8.66$\\
CPC \cite{kahn2019librilight} & $6.51$&$8.42$&$6.22$&$8.55$&$7.43$\\
\midrule
\multicolumn{6}{l}{\textit{Training set \textbf{NOT} in Libri-light}}\\
Topline & $4.36$&$6.16$&$4.22$&$5.87$&$5.15$\\
CPC+DA \cite{Kharitonov2020data_augment} & $4.66$&$5.81$&$4.46$&$6.56$&$5.37$\\
C-BNF-Du & $5.50$&$7.50$&$5.27$&$6.86$&$6.28$\\
C-BNF-Ma & $7.63$&$9.30$&$7.28$&$9.51$&$8.43$\\
\midrule[0.2pt]
\bottomrule
\end{tabular}%
}
\label{tab:exp_results_dnn_bnf}
\end{table}
Next, we compare the overall ABX error rates ($\%$) of the proposed approach, comparative approaches, and our created supervised topline on the Libri-light evaluation sets.  Table \ref{tab:exp_results_dnn_bnf} shows the performances of the different approaches evaluated on the four Libri-light evaluation sets separately and averaged over all evaluations sets (right-most column), for the across-speaker condition (top half of of Table \ref{tab:exp_results_dnn_bnf}) and within-speaker condition (bottom half of Table \ref{tab:exp_results_dnn_bnf}). 
The systems CPC \cite{kahn2019librilight} and   CPC+DA \cite{Kharitonov2020data_augment} adopt the CPC model, and CPC+DA additionally applies data augmentation. 
Since CPC+DA \cite{Kharitonov2020data_augment} used  augmented audio data derived from Librispeech, the performance of this system is listed under "Training set NOT in Libri-light".
The topline system,  C-BNF-Du and C-BNF-Ma that are trained with Librispeech, CGN and ADT respectively (see Table \ref{table:expsetup_naming}) are also listed under “NOT in Libri-light”.
Several observations can be made from this table:


(1) The cross-lingual phone-aware DNN-BNF methods that use the APC features as input features (A-BNF-Du and A-BNF-Ma) consistently outperform  systems with MFCC input features (M-BNF-Du and M-BNF-Ma) on all evaluation sets. This demonstrates the effectiveness of the front-end APC pretraining in our proposed two-stage system framework. This observation confirms our earlier findings in recent work  \cite{feng2020unsupervised}, in which the training set for APC was unlab-600 ($526$ hours).
The present study shows that when the amount of training material is scaled up to unlab-6K (5,273 hours), APC pretraining brings even greater relative ABX error rate reduction than when trained on unlab-600: the across- and within-speaker relative error rate reductions from M-BNF-Du to A-BNF-Du are $9.5\%$ and $5.5\%$, respectively, when trained with unlab-6K, while they are $7.6\%$ and $4.8\%$ when  trained with unlab-600. Similarly, the across- and within-speaker relative error rate reductions from M-BNF-Ma to A-BNF-Ma are
$15.0\%$ and $15.4\%$, respectively, when trained with unlab-6K, while $10.0\%$ and $9.3\%$ when trained with   unlab-600.


(2) The proposed A-BNF-Du system trained with unlab-600 is comparable to the state-of-the-art CPC+DA system \cite{Kharitonov2020data_augment}\footnote{A more strict comparison between A-BNF-Du and CPC+DA should be made under the identical training material setting, however performance of CPC+DA trained with Libri-light datasets was not reported in [65].}. When A-BNF-Du is trained with the larger, unlab-6K set, it outperforms CPC+DA. Both A-BNF-Du and A-BNF-Ma outperform CPC  \cite{kahn2019librilight}
trained on   unlab-600 and unlab-6K.
It should be noted that in contrast to our approach, the CPC and CPC+DA systems do not require transcribed data from OOD languages for training. The huge advancement of CPC+DA  \cite{Kharitonov2020data_augment} over CPC \cite{kahn2019librilight} indicates the effectiveness of adopting data augmentation techniques in the concerned task. It would thus be interesting to investigate the efficacy of integrating data augmentation and CPC with our current system framework. We leave it for future study.  


(3) The best performance achieved by our proposed approach (A-BNF-Du) trained on unlab-6K is only slightly inferior to the supervised topline system ($0.38\%$ across-speaker and $0.05\%$ within-speaker absolutely).  This is an encouraging finding, as it indicates that on the task of discriminating between  a pair of subword units of an unknown language,
with a sufficient amount of unlabeled data,  our approach which does not  require any linguistic knowledge of the target language, could perform on par with a supervised AM using transcribed training data of that language.

(4) In both the cross-lingual phone labeling method and the cross-lingual AM based BNF method, using Dutch data as OOD resources results in better performance than using Mandarin data. For A-BNF-Du and A-BNF-Ma that adopt cross-lingual phone labeling, the superiority of using Dutch data over using Mandarin data was found in a recent work \cite{feng2020unsupervised}. The present study demonstrates the same finding  in the cross-lingual AM based method by comparing  C-BNF-Du and C-BNF-Ma.

(5) With $600$-hour (or more) unlabeled training data of the target language available, the cross-lingual phone labeling method outperforms the cross-lingual AM based BNF method which does not rely on target unlabeled speech data but relies on OOD labeled speech data.  This can be observed by, for instance, comparing A-BNF-Du with C-BNF-Du, or by comparing A-BNF-Ma with C-BNF-Ma. The superiority of cross-lingual phone labeling is consistent over all the evaluation sets and both the across- and within-speaker conditions. This superiority can be partially explained as the first method leverages both OOD transcribed data and in-domain unlabeled data in system development, while the second method leverages OOD transcribed data only\footnote{We do not claim that   our method always outperforms the second one. In our unpublished results, A-BNF-Du trained with a 13-hour subset of unlab-600 performed worse than C-BNF-Du, which implies that large amounts of unlabeled target training data are required in order to get a good performance with our method.}.

Interestingly, the superiority of cross-lingual phone labeling over cross-lingual AM based BNF is more prominent when Mandarin data is chosen as OOD resource, compared to when Dutch data is chosen. For instance, in the across-speaker condition, the relative performance increase from C-BNF-Du to A-BNF-Du (trained with unlab-6K) is 
$14.1\%$ relatively,
while for the Mandarin models this relative increase is 
$30.7\%$.
A possible explanation is that the cross-lingual AM based BNF method has a language mismatch   between training  and test acoustic data,
while for the cross-lingual phone labeling method, the acoustic data during training and test are both from the target language. A larger language mismatch between the OOD language and the target language, e.g. Mandarin-English has a larger mismatch than Dutch-English, leads to larger negative effects on ABX performance.








\subsection{ZeroSpeech 2017 evaluation results}

\begin{table}[!t]
\renewcommand\arraystretch{0.50}
\centering
\caption{Overall ABX error rates ($\%$) of A-BNF-Du, A-BNF-Ma, the official MFCC baseline, the supervised topline, and state-of-the-art  systems  on the ZeroSpeech 2017  English evaluation sets. The topline and \textit{SH} that are trained with supervised English data are marked with $^\dagger$.}
\resizebox{1 \linewidth}{!}{%
\begin{tabular}{l|ccc|c|ccc|c}      
\toprule
\midrule[0.2pt]
& \multicolumn{4}{c}{Across-speaker} & \multicolumn{4}{c}{Within-speaker}\\
  & 1s & 10s & 120s &  Avg. & 1s & 10s & 120s &  Avg.\\ 
\midrule
\multicolumn{9}{l}{\textit{Training set: unlab-600}} \\
A-BNF-Du &$7.65$&$6.69$&$6.66$&$7.00$&$5.52$&$4.77$&$4.68$&$4.99$\\
A-BNF-Ma & $8.19$&$7.33$&$7.30$&$7.61$&$ 5.97$&$5.39$&$5.37$&$5.58$ \\
APC &$14.36$&$12.59$&$12.49$&$13.15$&$9.80$&$8.28$&$8.26$&$8.78$\\


\midrule
\multicolumn{9}{l}{\textit{Training set: \underline{unlab-6K}}} \\
A-BNF-Du &  $\bm{6.91}$&$6.28$&$\bm{6.26}$&$\underline{\bm{6.48}} $&$\bm{4.96}$&$\bm{4.53}$&$\bm{4.54}$&$\underline{\bm{4.68}}$\\
A-BNF-Ma &$7.47$&$6.93$&$6.89$&$7.10$&$5.38$&$5.03$&$5.00$&$5.14$ \\
APC & $12.00$&$10.79$&$10.70$&$11.16$&$8.32$&$7.28$&$7.21$&$7.60$\\

\midrule
\multicolumn{9}{l}{\textit{Training set: ZeroSpeech 2017 ($45$ hours)}}\\
CH \cite{chorowski2019unsupervised} & $8.1$&$8.0$&$8.0$&$8.03$&$5.6$&$5.5$&$5.5$&$5.53$\\
HE \cite{heck2017feature} & $10.1$&$8.7$&$8.5$&$9.10$&$6.9$&$6.2$&$6.0$&$6.37$\\
\midrule
\multicolumn{9}{l}{\textit{Training set \textbf{NOT} in Libri-light or ZeroSpeech 2017}}\\
Baseline \cite{dunbar2017zero} & $23.4$&$23.4$&$23.4$&$23.4$&$12.0$&$12.1$&$12.1$&$12.1$\\
$^{\dagger}$Topline \cite{dunbar2017zero} &$8.6$&$6.9 $&$6.7$&$7.40$&$6.5$&$5.3$&$5.1$&$5.63$\\
$^{\dagger}$SH \cite{shibata2017composite} & $7.9$&$7.4$&$6.9$&$7.40$&$5.5$&$5.2$&$4.9$&$5.20$\\
CPC+DA \cite{Kharitonov2020data_augment} & -& $\bm{5.8}$&- & -&-&$4.6$&-&- \\
C-BNF-Du &$7.81$&$7.79$&$7.78$&$7.79$&$5.59$&$5.58$&$5.60$&$5.59$\\
C-BNF-Ma &$10.60$&$10.62$&$10.57$&$10.60$&$7.67$&$7.58$&$7.56$&$7.60$\\
 
\midrule[0.2pt]
\bottomrule
\end{tabular}%

}
\label{tab:exp_results_zrsc2017}
\end{table}
Finally, we tested our approach on the ZeroSpeech 2017 data. Overall ABX error rates  ($\%$) of the proposed approach (A-BNF-Du and A-BNF-Ma), the cross-lingual AM based BNF approach (C-BNF-Du and C-BNF-Ma), and the front-end APC representation on the ZeroSpeech 2017 English 1s, 10s and 120s evaluation sets are listed in Table \ref{tab:exp_results_zrsc2017} separately and averaged over all evaluation sets. The official baseline and topline of ZeroSpeech 2017,   systems   \textit{HE} \cite{heck2017feature}, \textit{CH} \cite{chorowski2019unsupervised}, \textit{SH} \cite{shibata2017composite} that are at the top of the ZeroSpeech 2017 leaderboard  and the system \textit{CPC+DA} \cite{Kharitonov2020data_augment} (same as CPC+DA discussed in Section \ref{subsec:exp_results_back_end})  are listed in the table for comparison. The systems \textit{HE} and \textit{CH} utilized untranscribed speech data from the ZeroSpeech 2017 training sets only. The system \textit{SH} utilized over $1,300$ hours of OOD transcribed data including $80$ hours of English data during training, so \textit{SH} is a supervised system. Please note that only the results on the $10s$ evaluation set of the CPC+DA approach are reported in \cite{Kharitonov2020data_augment}.

From Table \ref{tab:exp_results_zrsc2017} it can be seen that when trained with unlab-600, A-BNF-Du outperforms the supervised topline system. When trained with the larger, unlab-6K set, both A-BNF-Du and A-BNF-Ma outperform the topline. A-BNF-Du and A-BNF-Ma perform better than the unsupervised systems \textit{CH, HE} and the supervised system \textit{SH} both when trained with unlab-600 and with unlab-6K.
The state-of-the-art system CPC+DA performs better than our best system (A-BNF-Du, trained with unlab-6K) in the across-speaker condition on the 10s evaluation set, while ours is better in the within-speaker condition. A-BNF-Du and A-BNF-Ma both perform better than APC in the unlab-600 and unlab-6K training data settings.

Table \ref{tab:exp_results_zrsc2017}   shows
A-BNF-Du and A-BNF-Ma representations  perform better than C-BNF-Du and C-BNF-Ma representations. It  further confirms our finding in the previous section that of the two cross-lingual knowledge transfer methods, the cross-lingual phone labeling method (in the back-end of A-BNF-Du and A-BNF-Ma) performs better. The superiority of the cross-lingual phone labeling method over the cross-lingual AM based BNF method is more prominent when Mandarin data is chosen as the OOD resource, compared to when Dutch data is chosen. This is again in line with the finding in the previous section.



\section{In-depth analyses  of the effectiveness of the proposed approach}
\label{sec:analysis_on_results}
\textcolor{black}{The in-depth analyses of the performance of our proposed approach is conducted at two broad levels: at the phoneme level, using the phoneme-level ABX error rate (see Section \ref{subsec:analysis_phoneme_level}); and at the AF level, using the pairwise and attribute-level ABX AF error rates (see Section \ref{subsec:analysis_af_level}). 
These analyses aim to uncover what phoneme and AF information is (not) captured by the learned representation from our proposed approach, which can be used to guide future research in further improving  the proposed approach.
}
\subsection{Setup}
The in-depth analyses of the performance of  our proposed approach are conducted on the dev-clean set from Libri-light. This set contains $2,703$ utterances summing up to $5.4$ hours. The total number of frames is
$1,934,785$.

The A-BNF-Du and A-BNF-Ma representations generated by our proposed approach trained using unlab-600  are chosen for the analyses. Moreover, the front-end APC features trained with unlab-600 and the official MFCC features are chosen for analyses in order to compare against A-BNF-Du and A-BNF-Ma.  

The phoneme-level analysis uses the $39$ English phonemes in the CMU Dictionary \cite{cmu}: these are 10 monophthongs, 5 diphthongs, and 24 consonants. Calculation of $p_{co} (\omega_i)$ (see Equation (\ref{eqt:p_co})) depends on the ground-truth English phoneme labels and the cross-lingual phone labels. The English true phoneme labels for dev-clean are obtained by carrying out a forced alignment using the English TDNN AM that is described in Section \ref{subsec:exp_setup_baseline_topline}. Please note that during calculation of $p_{co} (\omega_i)$ using Mandarin cross-lingual phone labels, tone symbols are removed, in order to  make a fair comparison of $p_{co} (\omega_i)$ between Dutch and Mandarin phone labels.


The AF-level analysis focuses on four AFs: MoA and PoA for consonants, and height and backness for vowels. In the analysis of each AF, t-SNE  \cite{maaten2008visualizing} is adopted to visualize the distribution of the learned representations with respect to the different attributes in an AF (e.g. fricative is an attribute in MoA). The number of speech frames per AF for the visualizations is around $2,400$ for all four AFs as a trade-off between computational complexity and avoiding sparsity of the visualization plots. The number of speech frames per AF attribute is equal for all attributes for the same AF, i.e., $500$ for MoA, $300$ for PoA, and $800$ for vowel height and backness. The speech frames are randomly chosen from dev-clean. 







\subsection{Phoneme-level analysis results}
\label{subsec:analysis_on_results_phoneme_level}
\begin{figure}[!t]
    \centering
    \includegraphics[width= \linewidth]{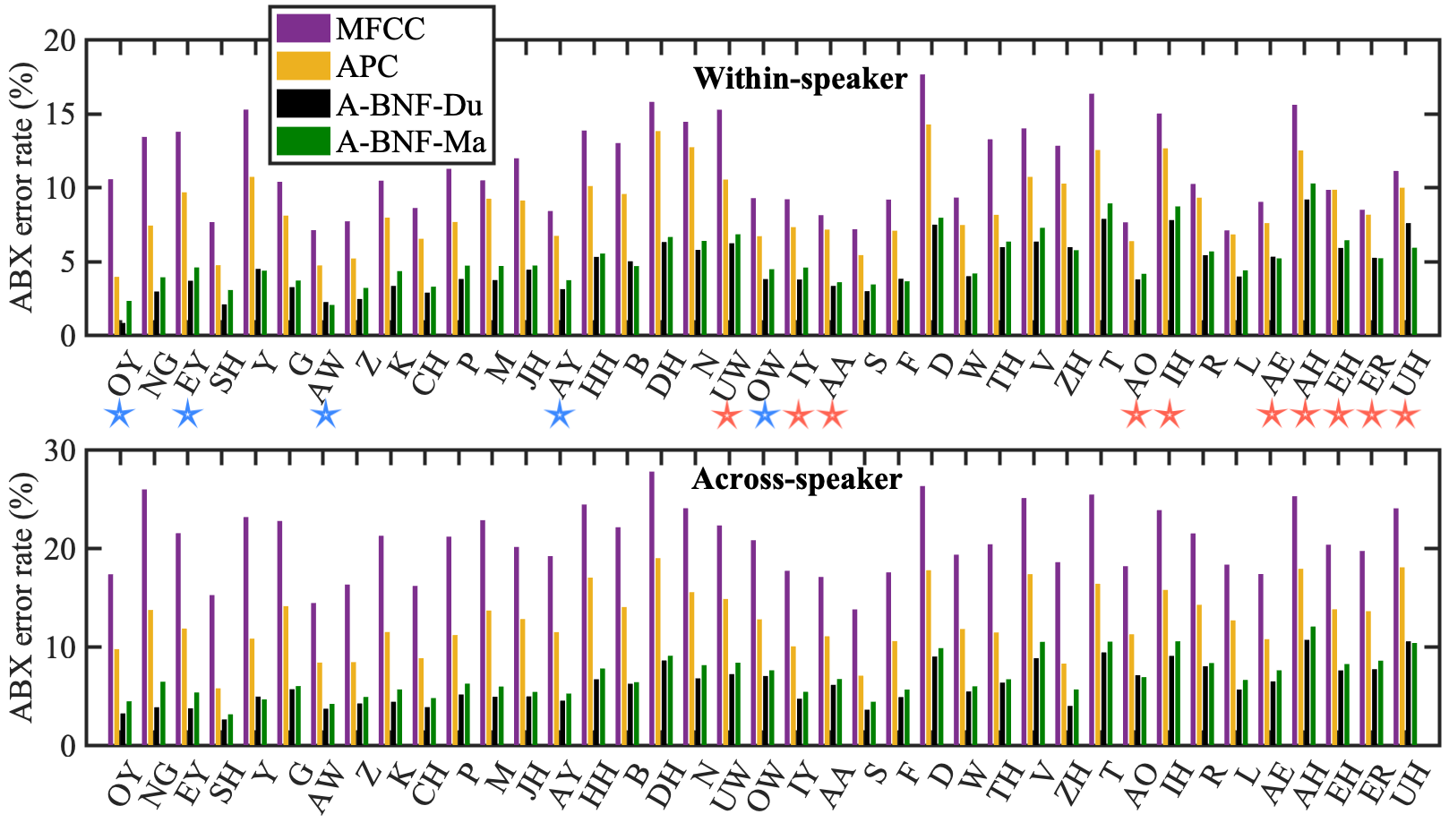}
    \caption{Phoneme-level ABX error rates ($\%$) of MFCC, APC, A-BNF-Du and A-BNF-Ma (lower is better). Phonemes are sorted (left to right) based on within-speaker relative error rate reduction from MFCC to A-BNF-Du in  descending order. Orange and blue  ``\ding{73}'' symbols denote monophthongs and diphthongs respectively, the rest are consonants. }
    \label{fig:analyses_per_phone_bar}
\end{figure}
\begin{figure}[!t]
    \centering
    \includegraphics[width= \linewidth]{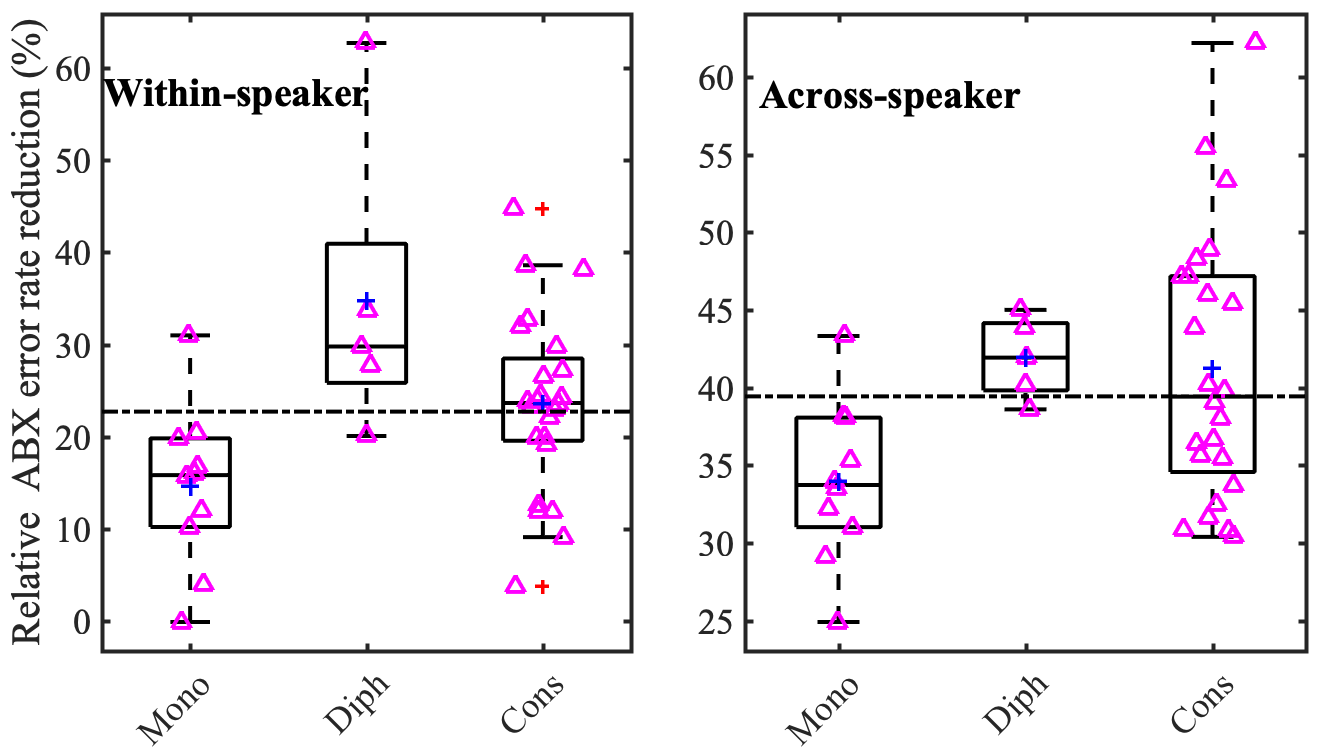}
    \caption{Distribution of phoneme-level relative ABX error rate reduction ($\%$) from MFCC to APC as the effect of front-end of the proposed approach (higher is better).  ``Mono'', ``Diph'' and ``Cons'' are abbreviations of Monophthongs, Diphthongs and Consonants.}
    \label{fig:analyses_per_phone_mfcc_to_apc_boxplot}
\end{figure}
\begin{figure}[!t]
    \centering
    \includegraphics[width= \linewidth]{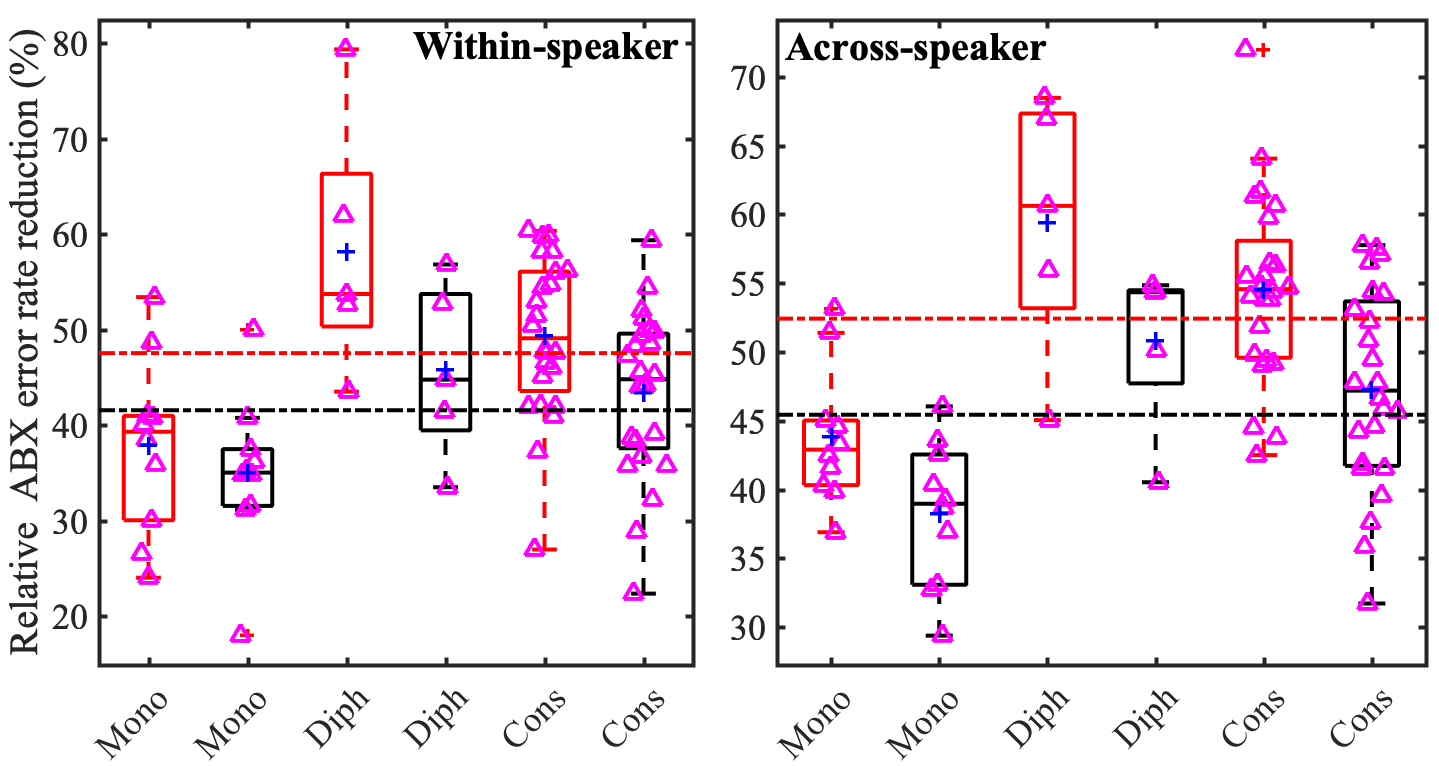}
    \caption{Distribution of phoneme-level relative ABX error rate reduction ($\%$) from APC to A-BNF-Du (red) and from APC to A-BNF-Ma (black) as the effect of back-end of the proposed approach (higher is better).}
    \label{fig:analyses_per_phone_apc_to_bnf_boxplot}
\end{figure}
\begin{figure}[!t]
    \centering
    \includegraphics[width= \linewidth]{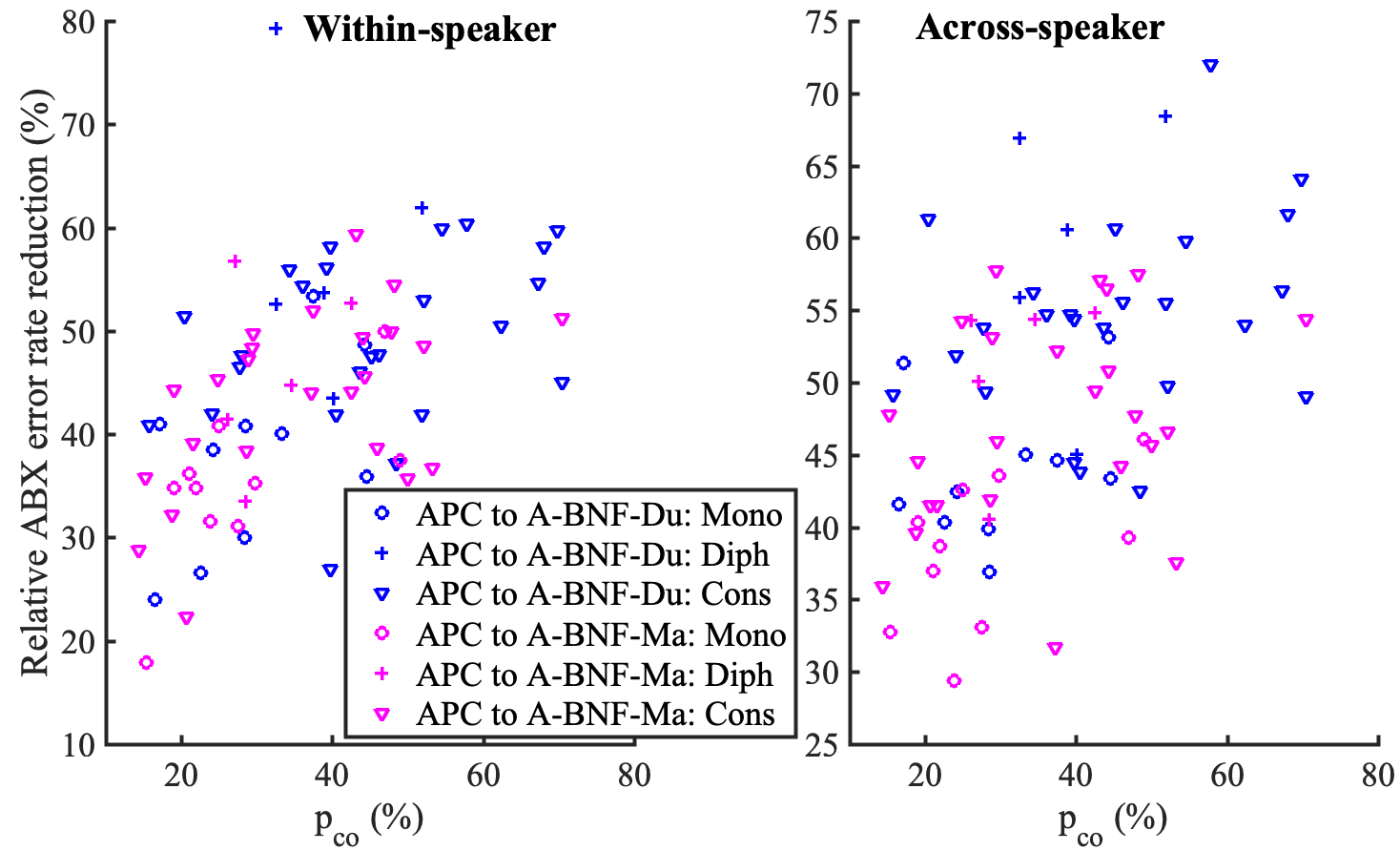}
    \caption{Correlation between phoneme-level ABX error rate reduction ($\%$) from APC to A-BNF-Du (blue)/A-BNF-Ma (pink) and $p_{co} (\%)$  for each English phoneme.}
    \label{fig:analyses_per_phone_apc_to_bnf_correlation}
\end{figure}

Phoneme-level ABX error rates (\%) of A-BNF-Du, A-
BNF-Ma, APC, and MFCC representations are illustrated in Fig.
 \ref{fig:analyses_per_phone_bar} for the within-speaker (top panel) and across-speaker (bottom panel) conditions separately. The phonemes are sorted (left to right) based on the within-speaker relative error rate reduction from MFCC to A-BNF-Du in descending order. The distribution of the phoneme-level relative ABX error rate reductions ($\%$) from
MFCC to APC feature representations aggregated for monophthongs, diphthongs, and consonants for the within-speaker (left panel) and across-speaker (right panel) conditions are shown in Fig. \ref{fig:analyses_per_phone_mfcc_to_apc_boxplot}, and from APC to A-BNF-Du (red boxes) and to A-BNF-Ma (black boxes) in Fig.  \ref{fig:analyses_per_phone_apc_to_bnf_boxplot}.
Figs. \ref{fig:analyses_per_phone_mfcc_to_apc_boxplot} and \ref{fig:analyses_per_phone_apc_to_bnf_boxplot} reflect the phoneme-level ABX error reduction achieved by the front-end and the back-end respectively. 
In Figs. \ref{fig:analyses_per_phone_mfcc_to_apc_boxplot} and \ref{fig:analyses_per_phone_apc_to_bnf_boxplot}, each triangle point represents an individual phoneme. The horizontal dash-dotted line in each subfigure of Figs.  \ref{fig:analyses_per_phone_mfcc_to_apc_boxplot} and \ref{fig:analyses_per_phone_apc_to_bnf_boxplot}
denotes the average of the phoneme-level relative error rate reduction over all phonemes. The ``+'' inside each box marks the mean value of all phonemes in that box. 

Fig. \ref{fig:analyses_per_phone_bar} shows that diphthongs /OY/, /EY/, /AW/ and consonants /NG/, /SH/, /Y/, /G/ benefit the most from the learned A-BNF-Du representation of all $39$ phonemes. In contrast, monophthongs /UH/, /ER/, /EH/, /AH/, /AE/ and consonants /R/ and /L/ benefit the least. Interestingly, phonemes with the largest error rate reductions (left-most in Fig. \ref{fig:analyses_per_phone_bar}) are not only those having the highest error rates by the MFCC representation
(e.g. /Y/ has a high error rate but /SH/ does not).
In general, performance improvements obtained from MFCC to A-BNF-Du are larger for diphthongs than for monophthongs. 
This can be clearly observed when comparing the orange (denote monophthongs) and blue (denote diphthongs) stars in Fig.  \ref{fig:analyses_per_phone_bar}.
For consonants, the improvements vary greatly. As is seen from Fig. \ref{fig:analyses_per_phone_bar}, some consonants (/NG/ and /SH/) are among the phonemes with the largest improvements, while some (/L/ and /R/) are among the least.


Fig. \ref{fig:analyses_per_phone_mfcc_to_apc_boxplot} shows that  with front-end APC pretraining, all phonemes show a
positive error rate reduction (except the monophthong vowel /EH/ in the within-speaker condition). 
This demonstrates APC is effective not only from the perspective of overall ABX performance, which is shown in the previous section, but also towards each individual phoneme. Nevertheless, the gains differ for the different phonemes: one diphthong /OY/ gains a huge improvement  (over $60\%$) when using APC  in the within-speaker condition. while the biggest improvements in the across-speaker condition were found for three consonants: over $50\%$ relative error rate reductions for /Y/, /SH/ and /ZH/. At the same time, the two monophthongs /EH/, /ER/ and consonant /L/ benefit little from APC pretraining in the within-speaker condition  (less than $5\%$). 

Fig. \ref{fig:analyses_per_phone_apc_to_bnf_boxplot} shows that irrespective of using Dutch or Mandarin labels as cross-lingual labels, diphthongs benefit the most from the back-end cross-lingual BNF learning, followed by consonants, while monophthongs benefit the least. Moreover,  using Dutch labels results in larger performance improvements than using Mandarin labels on all three phoneme categories. The advantage of the Dutch labels over the Mandarin ones is also illustrated in Fig. \ref{fig:analyses_per_phone_bar}, where the majority of the phoneme-level ABX error rates by A-BNF-Du are lower than those by A-BNF-Ma (/Y/ and /UH/ are two exceptions in both the within- and the across-speaker condition).

From Fig. \ref{fig:analyses_per_phone_mfcc_to_apc_boxplot} and Fig.  \ref{fig:analyses_per_phone_apc_to_bnf_boxplot}, 
it can be observed that both the front-end and the back-end of the proposed approach bring larger performance improvements for diphthongs than monophthongs. Recall in Fig. \ref{fig:analyses_per_phone_bar} we saw that A-BNF-Du  achieved larger performance improvements over MFCC for diphthongs than for monophthongs, we conclude that the superiority of modeling diphthongs over monophthongs by A-BNF-Du results from a combined effect of APC pretraining and cross-lingual phone-aware DNN-BNF model training. The larger reduction in error rates  for diphthongs than for monophthongs in the APC front-end can possibly be attributed to APC pretraining being better at modeling duration or at least longer sounds: diphthongs, which are long sounds, and long monophthongs benefit the most from APC pretraining in the within-speaker condition, i.e., /UW/ and /IY/ (not explicitly marked in Fig. \ref{fig:analyses_per_phone_mfcc_to_apc_boxplot}), are both long monophthong vowels, while short monophthong vowels such as /UH/, /IH/ and /EH/ are among the monophthongs that benefit the least. Moreover, consonants that benefit the most from APC in the within-speaker condition are /NG/, /TH/, /SH/, /Z/. None of them are \textit{stops}, which have a short time duration.



To gain deeper insights on why the back-end DNN-BNF model performs differently in capturing different target phonemes' information,
Fig. \ref{fig:analyses_per_phone_apc_to_bnf_correlation} plots the correlation between the phoneme-level ABX error rate reduction ($\%$) from APC to A-BNF-Du/-Ma and $p_{co}$ ($\%$) for English phonemes. In line with our expectation, a clear positive correlation for both the A-BNF-Du (blue) and the A-BNF-Ma (pink) representations and for both the within-speaker and the across-speaker conditions can be observed: a high $p_{co} (\omega)$ of an English phoneme $\omega$ indicates a high consistency of cross-lingual phone labels that are assigned to frames of phoneme $\omega$. This means that frames of the same sounds in the target language are consistently assigned the same English label (irrespective of whether it is the \textit{correct} English label), thus ensuring reliable frame label supervision for training the back-end DNN-BNF model to create a subword-discriminative feature representation  of $\omega$.

Furthermore, from Fig. \ref{fig:analyses_per_phone_apc_to_bnf_correlation},  we can see that using Dutch phone labels (blue marks) results in more English phonemes getting a high $p_{co}$ than Mandarin phone labels (pink marks). 
Specifically, in the case of using Mandarin labels, only for /B/  $p_{co}$ is larger than $55\%$ (see horizontal axis of either one sub-figure in Fig. \ref{fig:analyses_per_phone_apc_to_bnf_correlation}), while this is the case for $7$ English phonemes (/S/, /M/, /K/, /N/, /P/, /NG/, /G/) when using Dutch labels. 
Interestingly, all of these are consonants. For monophthongs and diphthongs, regardless of using Dutch or Mandarin labels, their respective $p_{co}$ rarely exceeds $50\%$ (/EY/ is the only exception when using Dutch labels).  This indicates that both monophthongs and diphthongs are less likely to obtain highly consistent cross-lingual phone labels than consonants, which explains why monophthongs benefit less from back-end cross-lingual DNN-BNF model training than consonants  (see Fig. \ref{fig:analyses_per_phone_apc_to_bnf_boxplot}): the frame label supervision for training the DNN-BNF model is of less quality for monophthongs than for consonants.

It is worth noting that diphthongs 
were substantially benefiting from the back-end DNN-BNF model (see Fig. \ref{fig:analyses_per_phone_apc_to_bnf_boxplot}). 
However,  as mentioned above, diphthongs are less likely to obtain highly consistent cross-lingual phone labels than consonants, which appears to contrary to Fig. \ref{fig:analyses_per_phone_apc_to_bnf_boxplot}.
Our explanation is as follows: $p_{co}$ does not take into account the dynamic characteristics of diphthongs. Specifically, when cross-lingual phone labels are being generated by an OOD ASR system, it might well be that the first half and second half of the speech frames of a diphthong are labeled as two different cross-lingual phones. This would result in a low $p_{co}$, but does little to no harm to training the back-end DNN-BNF model to learn the representation of such a diphthong, because   the DNN-BNF would learn to represent the diphthong as a sequence of two consecutive phones. An example is the diphthong /OY/ modeled by A-BNF-Du, which is marked as the top “+” in the left half of Fig. \ref{fig:analyses_per_phone_apc_to_bnf_correlation}: while /OY/ gains the largest within-speaker ABX error rate reduction ($79.4\%$), $p_{co}(\mathrm{/OY/})$ is moderate ($32.5\%$). Among all the speech frames of /OY/, the Dutch phone [o]\footnote{IPA symbol is [o\textlengthmark]. The vowel in \textit{oost} (English translation: \textit{east}).} occupies $32.5\%$ of the frame labels, the Dutch phone [j]\footnote{IPA symbol is [j]. The vowel in \textit{ja} (English translation: \textit{yes}).} occupies $21.8\%$, and the rest are labeled as other Dutch phones with smaller percentages.

\subsection{AF-level analysis results: Manner of articulation}

%

\begin{figure}[!t]
    \centering
    \includegraphics[width= \linewidth]{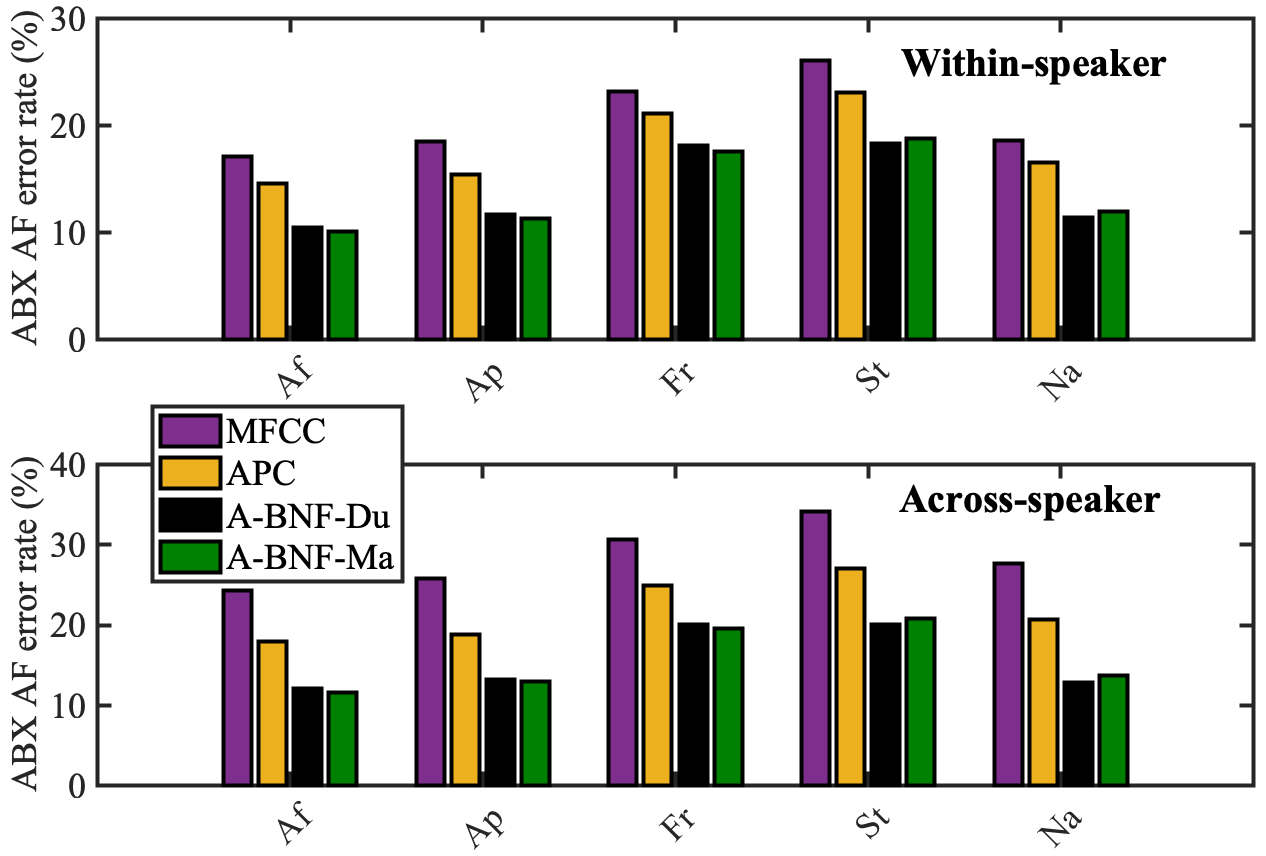}
    \caption{MoA attribute-level ABX AF error rate ($\%$) of MFCC, APC, A-BNF-Du, and A-BNF-Ma.}
    \label{fig:analyses_per_af_moa}
\end{figure}

Manner of articulation (MoA) attribute-level ABX AF error rates ($\%$) of the MFCC, APC, A-BNF-Du, and A-BNF-Ma representations are shown in Fig. \ref{fig:analyses_per_af_moa}. MoA pairwise ABX AF error rates ($\%$) of A-BNF-Du and A-BNF-Ma representations are listed in Table \ref{table:analyses_pairwise_af_moa}. 

\begin{table}[!t]
\centering
\caption{ MoA pairwise ABX AF error rates ($\%$) of A-BNF-Du and A-BNF-Ma.  \textcolor{magenta}{\textbf{Pink}} numbers denote across-speaker error rates, \textbf{black} numbers denote within-speaker error rates. }
\begin{minipage}{0.7\linewidth}
\subcaption{A-BNF-Du}
\resizebox{\linewidth}{!}{
\begin{tabular}{c|ccccc|c}
 &  Af & Ap & Fr & St & Na & avg.   \\ 
\midrule
 Af &-& $2.96$  & $19.69$  & $16.76$  & $2.43$ & \multirow{4}{*}{14.01} \\
 Ap &\textcolor{magenta}{ $3.98$} &- & $15.04$ & $14.28$ &$14.51$ \\
 Fr &\textcolor{magenta}{ $22.65$} &\textcolor{magenta}{ $16.44$} &- & $25.70$ & $12.04$\\
 St &\textcolor{magenta}{$18.69$} &\textcolor{magenta}{ $15.88$} &\textcolor{magenta}{$27.48$} &- & $16.70$ \\
 Na &\textcolor{magenta}{$3.05$} &\textcolor{magenta}{$16.59$} &\textcolor{magenta}{$13.47$} &\textcolor{magenta}{$18.15$} &-\\
 \midrule
 avg. & \multicolumn{4}{c}{\textcolor{magenta}{15.72}}& \\
\end{tabular}
}
\end{minipage}
\begin{minipage}{0.7\linewidth}
\subcaption{A-BNF-Ma}
\resizebox{\linewidth}{!}{
\begin{tabular}{c|ccccc|c}
 &  Af & Ap & Fr & St & Na  & avg.   \\ 
\midrule
 Af &-& $2.62$ & $18.54$ & $17.01$ &    $2.23$ & \multirow{4}{*}{13.96}\\
 Ap &\textcolor{magenta}{$3.26$} &- &  $14.21$ & $13.76$  & $14.74$ \\
 Fr &\textcolor{magenta}{$20.68$} &\textcolor{magenta}{$15.68$} &- & $25.61$  &$12.08$  \\
 St &\textcolor{magenta}{$19.22$} &\textcolor{magenta}{$15.36$} &\textcolor{magenta}{$28.13$} &- & $18.82$  \\
 Na &\textcolor{magenta}{$3.07$} &\textcolor{magenta}{$17.53$} &\textcolor{magenta}{$13.77$} &\textcolor{magenta}{$20.36$} &-\\
 \midrule
 avg. & \multicolumn{4}{c}{\textcolor{magenta}{15.64}} & 
\end{tabular}
}
\end{minipage}
 
\label{table:analyses_pairwise_af_moa}
\end{table}
\begin{figure}[!t]
    \begin{subfigure}{0.49\linewidth}
	   \centering
	   \includegraphics[width=1\linewidth]{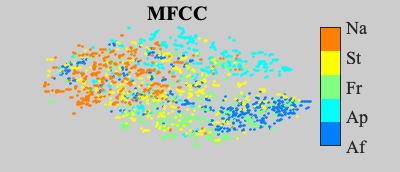}
    \end{subfigure}
   \begin{subfigure}{0.49\linewidth}
	   \centering
	   \includegraphics[width=1\linewidth]{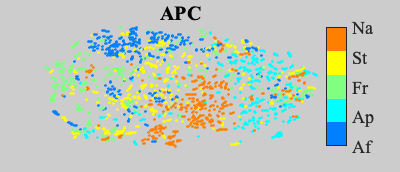}
    \end{subfigure}
    \newline
   \begin{subfigure}{0.49\linewidth}
	   \centering
	   \includegraphics[width=1\linewidth]{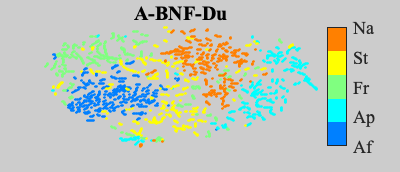}
    \end{subfigure}    \begin{subfigure}{0.49\linewidth}
	   \centering
	   \includegraphics[width=1\linewidth]{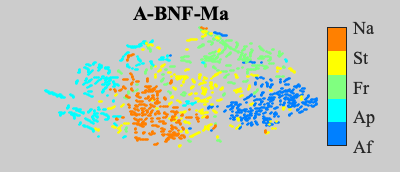}
    \end{subfigure}
    \caption{T-SNE visualization of the frame-level MFCC, APC, A-BNF-Du and A-BNF-Ma representations. Each color denotes a different MoA attribute.}
    \label{fig:analysis_tsne_moa}
\end{figure}

Fig. \ref{fig:analyses_per_af_moa} shows that both A-BNF-Du and A-BNF-Ma representations outperform MFCC and APC in capturing manner of articulation information. This shows that our approach proposed for modeling subword units implicitly learns information regarding  manner of articulation of phonemes. A-BNF-Du better captures stop and nasal information than A-BNF-Ma, while A-BNF-Ma better captures affricate, approximant and fricative information. The comparison on the average MoA pairwise ABX AF error rates for A-BNF-Du and A-BNF-Ma (right-most column and bottom row in Table \ref{table:analyses_pairwise_af_moa}) shows slightly lower average error rates for A-BNF-Ma than for A-BNF-Du in both the within-speaker and across-speaker conditions. This is in contrast to the phoneme-level results which showed consistently better results for A-BNF-Du over A-BNF-Ma. Taken together, this suggests that manner of articulation information is less language-dependent than phoneme information.

 Fig. \ref{fig:analyses_per_af_moa} furthermore shows that stops and fricatives have consistently higher ABX AF error rates than the other three MoA attributes for MFCC, APC, A-BNF-Du, and A-BNF-Ma representations.  Table  \ref{table:analyses_pairwise_af_moa}  shows that fricatives are often confused with affricates and stops while stops are often confused with fricatives. From an articulatory-acoustic point of view this can be explained by stops being very short and highly dynamic sounds. They start with a stretch of silence or low noise (in the case of voiced stops) when the vocal tract is fully closed, followed by a release that consists of noise caused by turbulence of the air in the vocal tract. A fricative solely consists of this noise, while an affricate can be seen as a concatenation of a short  stop and a short fricative, so has acoustic characteristics of both.

Fig. \ref{fig:analysis_tsne_moa} shows the t-SNE  visualizations  of the speech frames when using the MFCC,  APC, A-BNF-Du,  and  A-BNF-Ma  feature representations and labeling the frames with their MoA attributes. Each color represents a different MoA attribute, and each sample point  stands  for  a  speech  frame.  This figure clearly demonstrates that using A-BNF-Du and A-BNF-Ma as in our  proposed approach results in MoA attributes that form more explicit attribute-specific patterns than with MFCC and APC (compare bottom two panels with the top two panels). 

For A-BNF-Du and A-BNF-Ma, the clusters of affricates, approximants and nasals are more coherent than those of fricatives and stops. This is in agreement with the relatively higher ABX AF error rates of fricative and stop as shown in Fig. \ref{fig:analyses_per_af_moa}. Moreover, for the A-BNF-Du and A-BNF-Ma representations, affricates generally separate well from approximants and nasals, and are less separable from fricatives and from stops, which is in line with the articulatory-acoustic properties of affricates, fricatives, and stops. Table \ref{table:analyses_pairwise_af_moa} further confirms this finding: MoA pairwise error rates of affricate-fricative and affricate-stop are much higher  than those of affricate-approximant and affricate-nasal. By comparing visualizations of A-BNF-Du and A-BNF-Ma with MoA attributes, no noticeable difference is found.



\subsection{AF-level analysis results: Place of articulation}
\begin{figure}[!t]
    \centering
    \includegraphics[width= \linewidth]{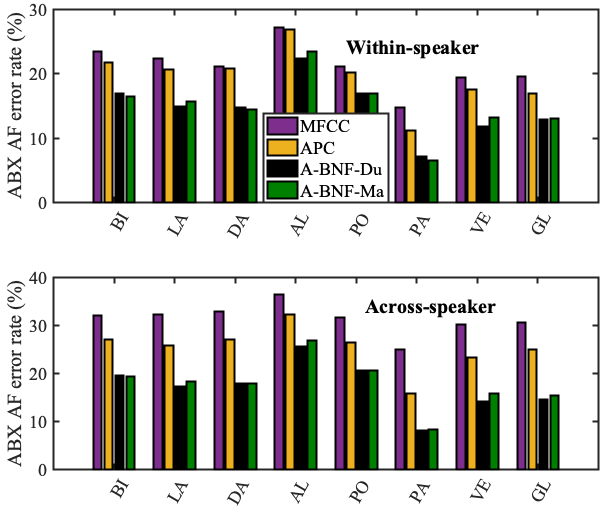}
    \caption{PoA attribute-level ABX AF error rate ($\%$) of MFCC, APC, A-BNF-Du and A-BNF-Ma.}
    \label{fig:analyses_per_af_poa}
\end{figure}

Place of articulation (PoA) attribute-level ABX AF error rates ($\%$) of MFCC, APC, A-BNF-Du, and A-BNF-Ma representations are shown in Fig. \ref{fig:analyses_per_af_poa}. PoA pairwise ABX AF error rates ($\%$) of A-BNF-Du and A-BNF-Ma representations are listed in Table \ref{table:analyses_pairwise_af_poa}.


\begin{table}[!t]
\caption{ PoA pairwise ABX AF error rates ($\%$) of A-BNF-Du and A-BNF-Ma.  \textcolor{magenta}{\textbf{Pink}} numbers denote across-speaker error rates, \textbf{black} numbers denote within-speaker error rates. }

 
\begin{minipage}{\linewidth}
\centering
\subcaption{A-BNF-Du}
\resizebox{  \linewidth}{!}{
\begin{tabular}{c|cccccccc|c}
 &  BI & LA & DA & AL & PO   & PA &VE &GL & avg.\\ 
\midrule
 BI &-&20.01 &18.58& 25.50& 20.60& 5.66& 15.72& 12.70 &\multirow{8}{*}{14.74} \\
 LA &\textcolor{magenta}{23.05} &-&16.72&25.93&18.46&3.34&10.45&9.86 \\
 DA &\textcolor{magenta}{21.94} &\textcolor{magenta}{20.81} &-&27.72 &14.42 &4.82 &9.19& 11.39\\
 AL &\textcolor{magenta}{28.45} &\textcolor{magenta}{29.09} &\textcolor{magenta}{32.66} & - & 23.97& 13.61& 21.17& 18.50\\
 PO & \textcolor{magenta}{23.60}&\textcolor{magenta}{21.29} &\textcolor{magenta}{19.90} & \textcolor{magenta}{28.11}& - & 10.39& 13.25& 17.70\\
 PA &\textcolor{magenta}{6.23} &\textcolor{magenta}{3.16} &\textcolor{magenta}{3.61} &\textcolor{magenta}{15.01} & \textcolor{magenta}{16.26}& - & 2.93& 9.77\\
 VE & \textcolor{magenta}{18.48}&\textcolor{magenta}{12.41} &\textcolor{magenta}{11.56} &\textcolor{magenta}{23.84} & \textcolor{magenta}{16.98}&\textcolor{magenta}{4.11} &  - & 10.50\\
 GL &\textcolor{magenta}{15.09} &\textcolor{magenta}{11.31} &\textcolor{magenta}{14.27} &\textcolor{magenta}{21.67} &\textcolor{magenta}{18.87} &\textcolor{magenta}{8.74} &\textcolor{magenta}{11.58} & -\\
 \midrule
 avg. & \multicolumn{7}{c}{\textcolor{magenta}{17.22}} & \\
\end{tabular}
}
\end{minipage} 

\begin{minipage}{\linewidth}
\centering
\subcaption{A-BNF-Ma}
\resizebox{  \linewidth}{!}{
\begin{tabular}{c|cccccccc|c}
 &  BI & LA & DA & AL & PO   & PA &VE &GL & avg.\\ 
\midrule
 BI &-&   20.23 & 17.96 & 25.25 & 18.48 & 4.14 & 16.48 & 12.40 & \multirow{7}{*}{14.98}\\
 LA &\textcolor{magenta}{23.15} &-& 17.52&27.77&17.93&	3.38&	12.52&	10.09 \\
 DA &\textcolor{magenta}{21.50} &\textcolor{magenta}{22.18} &-&27.22&	14.16&	3.13&	9.21&	11.90 \\
 AL &\textcolor{magenta}{28.47} &\textcolor{magenta}{31.33} &\textcolor{magenta}{31.88} & - & 24.83&	14.17&	24.51&	19.81 \\
 PO & \textcolor{magenta}{21.90}&\textcolor{magenta}{	21.29} &\textcolor{magenta}{18.48} & \textcolor{magenta}{29.36}& - & 10.99&	14.20&	17.74 \\
 PA &\textcolor{magenta}{5.70} &\textcolor{magenta}{2.54} &\textcolor{magenta}{4.83} &\textcolor{magenta}{16.44} & \textcolor{magenta}{15.49}& - & 3.37&	7.01 \\
 VE & \textcolor{magenta}{19.80}&\textcolor{magenta}{		15.32} &\textcolor{magenta}{12.67} &\textcolor{magenta}{27.24} & \textcolor{magenta}{17.89}&\textcolor{magenta}{3.55} &  - & 12.53 \\
 GL &\textcolor{magenta}{15.14} &\textcolor{magenta}{12.38} &\textcolor{magenta}{14.17} &\textcolor{magenta}{23.59} &\textcolor{magenta}{19.87} &\textcolor{magenta}{9.38} &\textcolor{magenta}{14.02} & -\\
 \midrule
 avg. & \multicolumn{7}{c}{\textcolor{magenta}{17.84}}& \\
\end{tabular}
}
\end{minipage}
 
\label{table:analyses_pairwise_af_poa}
\end{table}
\begin{figure}[!t]
    \begin{subfigure}{0.49\linewidth}
	   \centering
	   \includegraphics[width=1\linewidth]{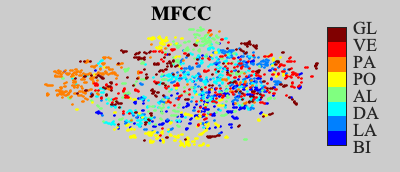}
    \end{subfigure}
   \begin{subfigure}{0.49\linewidth}
	   \centering
	   \includegraphics[width=1\linewidth]{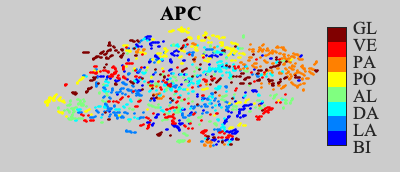}
    \end{subfigure}
    \newline
   \begin{subfigure}{0.49\linewidth}
	   \centering
	   \includegraphics[width=1\linewidth]{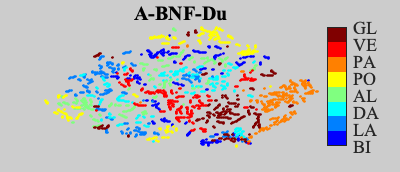}
    \end{subfigure}    \begin{subfigure}{0.49\linewidth}
	   \centering
	   \includegraphics[width=1\linewidth]{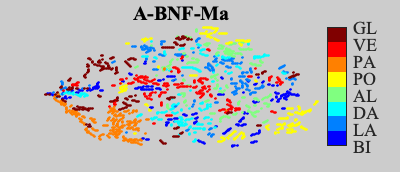}
    \end{subfigure}
    \caption{T-SNE visualization of the frame-level  MFCC, APC, A-BNF-Du and A-BNF-Ma representations. Each color denotes a different PoA attribute.}
    \label{fig:analysis_tsne_poa}

\end{figure}

Fig. \ref{fig:analyses_per_af_poa} shows that the A-BNF-Du and A-BNF-Ma representations consistently capture place of articulation information better than MFCC and APC for all PoA attributes and for both the within- and across speaker conditions. This demonstrates the effectiveness of our approach in capturing information that distinguishes PoA attributes. A-BNF-Du performs the best in capturing labiodental, alveolar and velar information in both within- and across-speaker conditions, while A-BNF-Ma performs the best in capturing bilabial, dental and palatal information in within-speaker conditions.
The comparison on the average PoA pairwise ABX AF error rates for A-BNF-Du and A-BNF-Ma (right-most column and bottom row in Table \ref{table:analyses_pairwise_af_poa}) shows little (less than absolute $0.25\%$) difference in both the within-speaker and across-speaker conditions. Moreover, Fig. \ref{fig:analyses_per_af_poa} shows no clear advantage between A-BNF-Du and A-BNF-Ma on PoA attribute-level ABX AF error rates.  This suggests that place of articulation information is less language-dependent than phoneme information, as the phoneme-level results showed consistently better results for A-BNF-Du over A-BNF-Ma.


Fig. \ref{fig:analyses_per_af_poa} shows that palatal has the lowest attribute-level ABX AF error rate. This is due to low pairwise ABX AF error rates between palatal and any other PoA attribute as shown in Table   \ref{table:analyses_pairwise_af_poa}. The fact that the A-BNF-Du and A-BNF-Ma are better able to capture palatal information than information of the other PoA attributes is likely due to the fact that palatal only concerns a single phoneme (/Y/) with a clear acoustic pattern. This is unlike the other PoA attribute that only consists of a single phoneme: glottal (/HH/). The attribute-level ABX AF error rate of glottal is much higher than that of palatal, and is similar to velar. This is likely due to the very low energy of /HH/ which makes it hard to distinguish the acoustics of /HH/ from silence and other phonemes (such as the first parts of stops).



Alveolar has the highest attribute-level ABX AF error rate for A-BNF-Du and A-BNF-Ma, which is due to high pairwise ABX AF error rates between alveolar and any other PoA attribute (see Table
 \ref{table:analyses_pairwise_af_poa}; except for the alveolar-palatal pair). Bilabial and postalveolar also have high attribute-level ABX AF error rates for A-BNF-Du and A-BNF-Ma: higher  than other attributes except alveolar. This is likely due to the PoA attributes of alveolar, bilabial and postalveolar containing phonemes with at least three different manners of articulation (see
 Table \ref{table:AF_moa_poa}), leading to highly diverse acoustics within each of the PoA attributes.


Fig. \ref{fig:analysis_tsne_poa} shows the t-SNE  visualizations  of the speech frames when using the MFCC,  APC, A-BNF-Du, and  A-BNF-Ma  feature representations and labeling the frames with their PoA attributes. Each color represents a different PoA attribute, and each sample point  stands  for  a  speech  frame. This figure clearly demonstrates that the PoA attributes form more explicit attribute-specific patterns when using the A-BNF-Du and A-BNF-Ma feature representations than with the MFCC and APC representations (compare the top panels with the bottom panels). The cluster of palatals is coherent in the  A-BNF-Du and A-BNF-Ma representations. The clusters of glottals and velars are coherent in A-BNF-Du, and less coherent in A-BNF-Ma. There are no coherent clusters of the other five PoA attributes shown in A-BNF-Du and A-BNF-Ma. This is consistent with Fig. \ref{fig:analyses_per_af_poa} which shows palatals, glottals and velars having lower attribute-level ABX AF error rates than the other five attributes in A-BNF-Du and A-BNF-Ma. Bilabials, labiodentals, and dentals show  overlap in both A-BNF-Du and A-BNF-Ma.

Comparing the results of the MoA and PoA analyses shows that the A-BNF-Du and A-BNF-Ma representations are better able to capture the underlying MoA and PoA information than the MFCC and APC representations. At the same time MoA information is better captured than PoA information by the A-BNF-Du and A-BNF-Ma representations (compare average pairwise ABX AF error rates in Tables \ref{table:analyses_pairwise_af_moa} and \ref{table:analyses_pairwise_af_poa}). This is in line with results found in \cite{ScharenborgGLM19representations} which showed that a naive feed-forward DNN trained for the vowel-consonant classification task captures manner of articulation information better than place of articulation information (without being explicitly trained to do so).


\subsection{AF-level analysis results: Vowel height and backness}
\begin{figure}[!t]
    \centering
    \includegraphics[width = \linewidth]{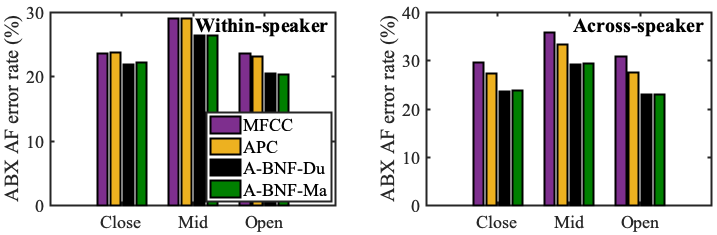}
    \caption{Vowel height attribute-level ABX AF error rate ($\%$) of MFCC, APC, A-BNF-Du and A-BNF-Ma.}
    \label{fig:analyses_per_af_vowel_height}
\end{figure}

\begin{figure}[!t]
    \centering
    \includegraphics[width = \linewidth]{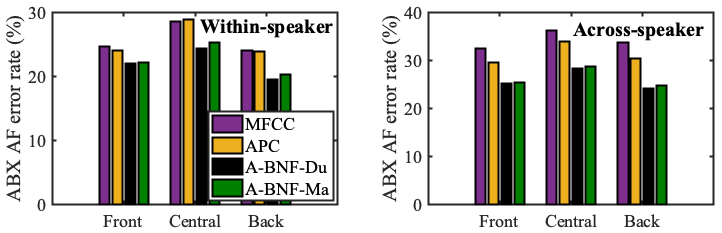}
    \caption{Vowel backness attribute-level ABX AF error rate ($\%$) of MFCC, APC, A-BNF-Du and A-BNF-Ma.}
    \label{fig:analyses_per_af_vowel_backness}
\end{figure}

Our final analyses focus on the effectiveness of our approach in capturing vowel height and backness information. Attribute-level ABX AF error rates ($\%$) for vowel height are illustrated in Fig. \ref{fig:analyses_per_af_vowel_height}, and those for vowel backness are illustrated in Fig. \ref{fig:analyses_per_af_vowel_backness}. Pairwise ABX AF error rates ($\%$) of A-BNF-Du and A-BNF-Ma are listed in Tables \ref{table:analyses_pairwise_af_height} and \ref{table:analyses_pairwise_af_backness}, respectively.

\begin{table}[!t]
\centering
\caption{ Vowel  height pairwise ABX AF error rates ($\%$) of A-BNF-Du and A-BNF-Ma.  \textcolor{magenta}{\textbf{Pink}} numbers denote across-speaker error rates, \textbf{black} numbers denote within-speaker error rates. ``Cl, Mi, Op'' stand for ``Close, Mid, Open''. }
\begin{minipage}{0.49\linewidth}
\subcaption{A-BNF-Du}
\resizebox{\linewidth}{!}{
\begin{tabular}{c|ccc|c}
 & Cl  & Mi & Op& avg.  \\ 
\midrule
 Cl &-&  27.75 & 15.99 & \multirow{2}{*}{22.93}\\
 Mi &\textcolor{magenta}{30.00} &-&25.06  \\
 Op &\textcolor{magenta}{17.38} & \textcolor{magenta}{28.47}&- \\
 \midrule
 avg. & \multicolumn{2}{c}{\textcolor{magenta}{25.40}} & \\
\end{tabular}
}
\end{minipage}
\begin{minipage}{0.49\linewidth}
\subcaption{A-BNF-Ma}
\resizebox{\linewidth}{!}{
\begin{tabular}{c|ccc|c}
 & Cl  & Mi & Op& avg.  \\ 
\midrule
 Cl &-&  28.28  & 16.08  & \multirow{2}{*}{23.02}\\
 Mi &\textcolor{magenta}{30.29} &-& 24.71 \\
 Op &\textcolor{magenta}{17.30} & \textcolor{magenta}{28.53}&- \\
 \midrule
 avg. & \multicolumn{2}{c}{\textcolor{magenta}{25.40}} & \\
\end{tabular}
}
\end{minipage}
 
\label{table:analyses_pairwise_af_height}
\end{table}
\begin{table}[!t]
\centering
\caption{ Vowel  backness pairwise ABX AF error rates ($\%$) of A-BNF-Du and A-BNF-Ma.  \textcolor{magenta}{\textbf{Pink}} numbers denote across-speaker error rates, \textbf{black} numbers denote within-speaker error rates. ``Fr, Ce, Ba'' stand for ``Front, Central, Back''. }
\begin{minipage}{0.49\linewidth}
\subcaption{A-BNF-Du}
\resizebox{\linewidth}{!}{
\begin{tabular}{c|ccc|c}
 & Fr  & Ce & Ba& avg.  \\ 
\midrule
 Fr &-& 26.78  & 17.20 & \multirow{2}{*}{21.98}\\
 Ce &\textcolor{magenta}{29.28} &-& 21.96 \\
 Ba &\textcolor{magenta}{21.12} & \textcolor{magenta}{27.35}&- \\
 \midrule
 avg. & \multicolumn{2}{c}{\textcolor{magenta}{25.90}} & \\
\end{tabular}
}
\end{minipage}
\begin{minipage}{0.49\linewidth}
\subcaption{A-BNF-Ma}
\resizebox{\linewidth}{!}{
\begin{tabular}{c|ccc|c}
 & Fr  & Ce & Ba& avg.  \\ 
\midrule
 Fr &-& 27.15 &17.34 & \multirow{2}{*}{22.64}\\
 Ce &\textcolor{magenta}{29.52} &-& 23.44 \\
 Ba &\textcolor{magenta}{21.47} & \textcolor{magenta}{27.94}&- \\
 \midrule
 avg. & \multicolumn{2}{c}{\textcolor{magenta}{26.25}} & \\
\end{tabular}
}
\end{minipage}
 
\label{table:analyses_pairwise_af_backness}
\end{table}

\begin{figure}[!t]
    \begin{subfigure}{0.495\linewidth}
	   \centering
	   \includegraphics[width=1\linewidth]{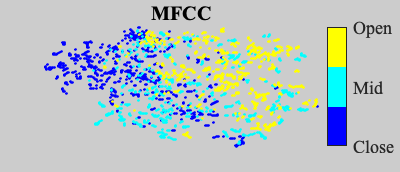}
    \end{subfigure}
   \begin{subfigure}{0.495\linewidth}
	   \centering
	   \includegraphics[width=1\linewidth]{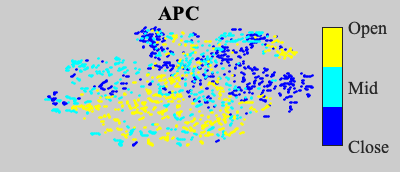}
    \end{subfigure}
    \newline
   \begin{subfigure}{0.495\linewidth}
	   \centering
	   \includegraphics[width=1\linewidth]{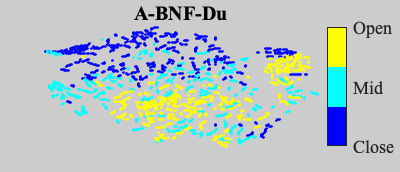}
    \end{subfigure}    \begin{subfigure}{0.495\linewidth}
	   \centering
	   \includegraphics[width=1\linewidth]{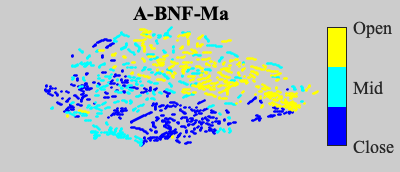}
    \end{subfigure}
    \caption{T-SNE visualization of the frame-level MFCC, APC, A-BNF-Du and A-BNF-Ma representations. Each color denotes a different vowel height attribute.}
        \label{fig:analysis_tsne_vowel_height}
\end{figure}
\begin{figure}[!t]
    \begin{subfigure}{0.495\linewidth}
	   \centering
	   \includegraphics[width=1\linewidth]{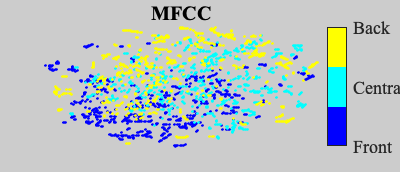}
    \end{subfigure}
   \begin{subfigure}{0.495\linewidth}
	   \centering
	   \includegraphics[width=1\linewidth]{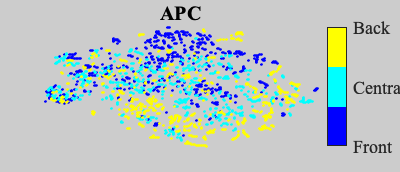}
    \end{subfigure}
    \newline
   \begin{subfigure}{0.495\linewidth}
	   \centering
	   \includegraphics[width=1\linewidth]{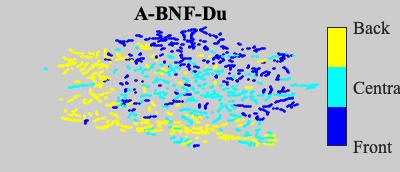}
    \end{subfigure}    \begin{subfigure}{0.495\linewidth}
	   \centering
	   \includegraphics[width=1\linewidth]{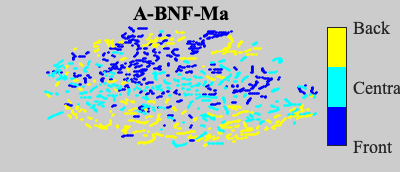}
    \end{subfigure}
    \caption{T-SNE visualization of the frame-level  MFCC, APC, A-BNF-Du and A-BNF-Ma representations. Each color denotes a different vowel backness attribute.}
        \label{fig:analysis_tsne_vowel_backness}

\end{figure}
Fig. \ref{fig:analyses_per_af_vowel_height} and Fig. \ref{fig:analyses_per_af_vowel_backness} show that, for both vowel height and backness, the attribute in the middle, i.e., \textit{mid} in height and \textit{central} in backness, performs consistently worse than the other attributes in both the within- and the across-speaker conditions. This worse performance is due to the  large confusion of \textit{mid} with \textit{close} and \textit{open} (see Table \ref{table:analyses_pairwise_af_height}), and of \textit{central} with \textit{front} and \textit{back} (see Table \ref{table:analyses_pairwise_af_backness}).

Fig. \ref{fig:analyses_per_af_vowel_height} and Fig. \ref{fig:analyses_per_af_vowel_backness} also show that,  
similar to what was observed for MoA and PoA for consonants, A-BNF-Du and A-BNF-Ma outperform MFCC and APC features in capturing the attributes of vowel height and backness. Comparing these results to the attribute-level ABX AF error rates for MoA (Fig. \ref{fig:analyses_per_af_moa}) and PoA (Fig. \ref{fig:analyses_per_af_poa}) shows that the monophthong vowel-related\footnote{Diphthongs are excluded in the vowel height and backness analyses, see Section \ref{subsec:analysis_af_level}.} AF attributes achieved higher error rates than the consonant-related ones  (i.e., MoA and PoA). In fact, most attribute-level ABX AF error rates for vowel height and backness obtained from A-BNF-Du and A-BNF-Ma fall within the $20\%-30\%$ range, whereas most of those of MoA and  PoA  fall  within the $10\%-20\%$ range.  This  observation  is  in accordance with findings in the phoneme-level analysis, where monophthongs are found to benefit less than consonants by our proposed approach (see Section \ref{subsec:analysis_on_results_phoneme_level}).

Fig. \ref{fig:analysis_tsne_vowel_height} and Fig. \ref{fig:analysis_tsne_vowel_backness} show the t-SNE visualizations of the frames when using the MFCC, APC, A-BNF-Du, and A-BNF-Ma feature representations and labeling them with vowel height and vowel backness respectively. Each color represents a different attribute in vowel height or backness, and each sample point stands for a speech frame.
Both figures show that the A-BNF-Du and A-BNF-Ma representations are better than MFCC and APC in capturing information that distinguishes vowel height and backness attributes, as they form more explicit attribute-specific patterns than with MFCC and APC.
Fig. \ref{fig:analysis_tsne_vowel_height} shows that for A-BNF-Du and A-BNF-Ma, while there are noticeable overlaps between the \textit{mid-close} pair and between the \textit{mid-open} pair, little overlap is observed between the \textit{close-open} pair.  It indicates that the information to distinguish the \textit{close} and \textit{open} attributes in vowel height are well learned by the proposed approach. This is in contrast to vowel backness (see Fig. \ref{fig:analysis_tsne_vowel_backness}) where there is extensive overlap between \textit{front},  \textit{central} and \textit{back} for A-BNF-Du and A-BNF-Ma.


Interestingly, the improvement of A-BNF-Du and A-BNF-Ma over MFCC in capturing vowel height and backness information seems not to be due to front-end APC pretraining.  Comparing APC with MFCC in Fig. \ref{fig:analyses_per_af_vowel_height} and Fig. \ref{fig:analyses_per_af_vowel_backness} shows that front-end APC pretraining has very limited or even a negative effect on capturing vowel height and backness information, especially in the within-speaker evaluation condition. This is contrary to what was observed for consonant information, where APC pretraining was found to be effective in capturing MoA and PoA (see Fig. \ref{fig:analyses_per_af_moa} and Fig. \ref{fig:analyses_per_af_poa}). These results are in line with those observed in the  phoneme analyses in Fig. \ref{fig:analyses_per_phone_mfcc_to_apc_boxplot}, where the efficacy of APC pretraining is more prominent on consonants than on monophthongs.




\section{Conclusion}
\label{sec:conclusion}
The present study addresses unsupervised subword modeling. A two-stage learning framework that consists of an APC front-end and a cross-lingual DNN-BNF back-end was proposed to tackle this problem. To evaluate the proposed approach, in addition to the widely adopted ABX subword discriminability metric, a comprehensive and systematic analysis was carried out at the phoneme-level and the articulatory feature (AF)-level to investigate the type of information that is (and is not) captured by the newly created feature representations. In order to do so, new metrics that focus on phoneme-level ABX subword discriminability and attribute-level ABX AF discriminability have been proposed.


Experiments were conducted using two databases: Libri-light and ZeroSpeech 2017. Using the overall ABX subword discriminability metric, 
the experimental results show that our approach is competitive or even superior to the state-of-the-art \cite{Kharitonov2020data_augment}. Front-end APC pretraining brings performance improvement to the entire learning framework compared to a system with only the DNN-BNF back-end. Performance further increased when the amount of training material was increased from $600$ hours to $6,000$ hours. The proposed system’s best performance, achieved by using $6,000$ hours of untranscribed training data without any linguistic knowledge of the target language, is very close to that of a supervised system trained on $1,000$-hour   transcribed data of the target language. Moreover,  the proposed back-end performs better than a cross-lingual AM based BNF method in exploiting cross-lingual knowledge transfer. 


Subsequent in-depth analyses investigated what information was captured by the newly created feature representations and this was compared to the information captured by baseline MFCC features and front-end APC features. The phoneme-level analysis showed that compared to MFCC, 
our two-stage approach achieves larger improvement in capturing diphthong information than monophthong vowel information,
and this is true in both the front-end and the back-end of our approach. For consonants, the improvement in capturing phoneme information from MFCC to our approach varies greatly to different consonants. Our results showed a positive correlation between the effectiveness of the back-end in capturing a target phoneme's information and the quality of cross-lingual phone labels assigned to that target phoneme.

The  AF-level analyses showed that the proposed approach is better than MFCC and front-end APC features in capturing manner and place of articulation information and vowel height and backness information. In the analysis of MoA, stop and fricative information are less well captured than affricate, approximant, and nasal information. The analysis of PoA showed that palatal is the best captured attribute, which is partially explained by the palatal AF attribute only consisting of a single phoneme /Y/, while most other PoA attributes consist of multiple phonemes with multiple manners of articulation. The analyses indicate MoA is better captured by the proposed approach than PoA which is in line with previous research \cite{ScharenborgGLM19representations}, and both MoA and PoA information are better captured than vowel height and backness information. Comparing the outcomes of the analyses at the AF and phoneme level suggests that AF information is less language-dependent than phoneme information, which is in line with the linguistic principles underlying articulatory features and phonemes.

In conclusion, both the front-end and back-end of the proposed approach are effective in capturing information that distinguishes individual phonemes. It demonstrates the importance of both the front-end and the back-end of our approach in the task of unsupervised subword modeling.  Regarding AF information, the front-end is effective in capturing MoA and PoA information, but is less well able to capture vowel height and backness information. In contrast, the back-end is effective in capturing all the MoA, PoA, vowel height and backness information. The phoneme-level and the AF-level analyses both indicate monophthong vowel information is much more difficult to capture than consonant information. This suggests that a possible direction to improve unsupervised subword modeling is investigating methods that improve the effectiveness of capturing monophthong vowel information. 

\bibliographystyle{IEEEtran}
\bibliography{mybib}
\clearpage
\renewcommand{\thepage}{S\arabic{page}} 
\renewcommand{\thesection}{S\arabic{section}}  
\renewcommand{\theequation}{S.\arabic{equation}}
\renewcommand{\thetable}{S\arabic{table}}  
\renewcommand{\thefigure}{S\arabic{figure}}

\beginsupplement
\section{Supplementary Material}
\label{sec:supple}
\subsection{Description of FHVAE}
\label{subsec:supple_fhvae_dtails}

We follow terminologies used in \cite{hsu2017nips} to describe details of the FHVAE model. Let $\mathcal{D}=\{\bm{X^{i}}\}_{i=1}^{M}$ denote a speech dataset with $M$ sequences. 
The $i$-th sequence $\bm{X^i}$ contains $N^i$ speech segments $\{\bm{x^{(i,n)}}\}^{N^i}_{n=1}$, where $\bm{x^{(i,n)}}$ is a segment of a fixed number of frames.
The FHVAE model formulates the generation process of a sequence $\bm{X}$ as\footnote{For simplicity, the superscript $i$ in $\bm{X^i}$ and  subsequent equations is omitted. This does not cause confusion.}  \cite{hsu2017nips},
\begin{enumerate}
    \item A vector $\bm{\mu_2 }$ is drawn from a prior distribution $p_{\theta}(\bm{\mu_2})=\mathcal{N} (\bm{0},\sigma^2_{\bm{\mu_2}} \bm{I})$;
    \item Latent segment variables $\bm{z_1 ^{n}} $ and latent sequence variables $\bm{z_2^{n}} $ are drawn from $p_{\theta}(\bm{z_1 ^{n}})=\mathcal{N} (\bm{0}, {\sigma^2_{\bm{z_1}}} \bm{I})$ and  $p_{\theta}(\bm{z_2 ^{n}| \bm{\mu_2}})=\mathcal{N}(\bm{\mu_2}, {\sigma^2_{\bm{z_2}}} \bm{I} )$;
    \item Speech segment $\bm{x^{n}}$ is drawn from \begin{equation}
    p_{\theta}(\bm{x^{n}}|\bm{z_1 ^{n}, \bm{z_2^{n}}})=\mathcal{N}(f_{\bm{\mu_x}} (\bm{z_1 ^{n}}, \bm{z_2^{n}}), diag(f_{\bm{\sigma^2_x}} (\bm{z_1 ^{n}}, \bm{z_2^{n}})). \label{eqt:supple_fhvae_generation_x}
        \end{equation}
\end{enumerate}

Here $\mathcal{N}$ denotes standard normal distribution, $ f_{\bm{\mu_x}} (\cdot, \cdot)$ and $ f_{\bm{\sigma^2_x}} (\cdot, \cdot)$ are parameterized by two DNNs.
Based on Equation (\ref{eqt:supple_fhvae_generation_x}), the joint probability for generating $\bm{X}$ is formulated as,
\begin{equation}
    p_{\theta} (\bm{\mu_2})\prod_{n=1}^{N} p_{\theta} (\bm{z_1^n}) p_{\theta} (\bm{z_2^{n}}|\bm{\mu_2})p_{\theta} (\bm{x^n}|\bm{z_1 ^{n}, \bm{z_2^{n}}}).
\end{equation}

The FHVAE introduces an inference model
to approximate the true posterior as follows,
\begin{equation}
   p_{\phi} (\bm{\mu_2})\prod_{n=1}^{N}p_{\phi} (\bm{z_2^n}| \bm{x^n}) p_{\phi}(\bm{z_1^n}|\bm{x^n}, \bm{z_2^n}).
   \label{eqt:inference}
\end{equation}
Here  $p_{\phi} (\bm{\mu_2}), p_{\phi} (\bm{z_2^n}| \bm{x^n})$ and $p_{\phi}(\bm{z_1^n}|\bm{x^n}, \bm{z_2^n})$ are all diagonal Gaussian distributions. 
The mean and variance values of $p_{\phi} (\bm{z_2^n}| \bm{x^n})$ and $p_{\phi}(\bm{z_1^n}|\bm{x^n}, \bm{z_2^n})$ are parameterized by   DNNs. 

The FHVAE optimizes the  \textit{discriminative segmental variational lower bound} $\mathcal{L} (\theta, \phi; \bm{x^{(i,n)}})$ \cite{hsu2017nips}, which is  defined as,
\begin{equation}
\begin{split}
&\mathbb{E}_{q_{\phi} (\bm{z_1^{(i,n)}},\bm{z_2^{(i,n)}}|\bm{x^{(i,n)}} )} [\log p_{\theta} (\bm{x^{(i,n)}}|\bm{z_1^{(i,n)}}, \bm{z_2^{(i,n)}})]\\
&-\mathbb{E}_{q_{\phi}(\bm{z_2^{(i,n)}} |\bm{x^{(i,n)}} )} [\mathrm{KL} (q_{\phi} (\bm{z_1^{(i,n)}}|\bm{x^{(i,n)}}, \bm{z_2^{(i,n)}})||p_{\theta} (\bm{z_1^{(i,n)}}))]\\
&-\mathrm{KL} (q_{\phi}(\bm{z_2 ^{(i,n)}}|\bm{x^{(i,n)}})|| p_{\theta}(\bm{z_2 ^{(i,n)}}| \bm{\tilde{\mu}_2^i})) \\
&+\frac{1}{N^i}\log p_{\theta} (\bm{\tilde{\mu}_2^i}) + \alpha \log p(i| \bm{z_2^{(i,n)}}),
\end{split}
\label{eqt:supple_fhvae_obj_function}
\end{equation}
where $\bm{\tilde{\mu}_2^{i}}$ denotes posterior mean of $\bm{\mu_2}$ for the $i$-th sequence, $\alpha$ denotes the  discriminative weight. The discriminative objective $\log p(i| \bm{z_2^{(i,n)}}) $ is formulated as,
\begin{equation}
    \log p(i| \bm{z_2^{(i,n)}}) := \log p_{\theta} (\bm{z_2^{(i,n)}}| \bm{\tilde{\mu}_2^i})-\log \sum_{j=1}^{M}p_{\theta} (\bm{z_2^{(j,n)}}| \bm{\tilde{\mu}_2^j}).
\end{equation}

After FHVAE training, 
$\bm{z_1}$ representation  is extracted as the desired speaker-invariant  representation of speech.

\subsection{Finding optimal parameters of APC models}
\label{subsec:supple_apc_para}
We report the experimental results on finding the optimal LSTM layer number and prediction step $n$ of the APC model.
Table \ref{tab:supple_exp_results_apc_paras} lists overall across-speaker and within-speaker ABX error rates ($\%$) of the APC features with different choices of the number of layers (left-most column) and prediction step $n$ (second column from left) on the four Libri-light evaluation sets separately and averaged over all evaluation sets (right-most column). All the models in this table are trained with the unlab-600 set.

Table \ref{tab:supple_exp_results_apc_paras} shows that  the optimal parameters of layer number and $n$ by considering both the across-speaker and within-speaker performances are 5 and 5 (see final column).

\begin{table}[!t]
\renewcommand\arraystretch{0.60}
\centering
\caption{Overall ABX error rates ($\%$) of APC features w.r.t different layer numbers and prediction steps $n$ on Libri-light. Bold indicates the best result  within a fixed number of layers. The symbol $^{\ddagger}$ indicates the optimal parameters of layer number and $n$ by considering both across- and within-speaker performances.}
\resizebox{1 \linewidth}{!}{%
\begin{tabular}{c|c|cccc|c}      
\toprule
\midrule[0.2pt]
\multicolumn{7}{c}{Across-speaker}\\
 \#layers & $n$ & dev-clean & dev-other & test-clean & test-other & Avg.\\ 
\midrule

\multicolumn{7}{l}{APC (\textit{training set: unlab-600})}\\

\multirow{5}{*}{3} 
  & 1 &$14.70$&$21.25$&$13.86$&$21.03$&$17.71$\\
  & 2 &$13.96$&$20.21$&$13.19$&$20.17$&$16.88$\\
  & 3 &$13.65$&$19.67$&$12.99$&$19.76$&$16.52$\\
  & 4 &$13.34$&$19.63$&$12.63$&$19.59$&$16.30$\\
  & 5 &$\bm{13.10}$&$\bm{19.50}$&$\bm{12.48}$&$\bm{19.42}$&$\bm{16.13}$\\

\midrule
\multirow{5}{*}{4} 
& 1 & $14.63$&$21.51$&$13.77$&$21.13$&$17.76$ \\
  & 2 & $14.07$&$20.25$&$13.32$&$20.23$&$16.97$
\\
  & 3 & $\bm{13.09}$&$\bm{19.14}$&$12.58$&$\bm{19.21}$&$\bm{16.01}$
\\
  & 4 & $13.20$&$19.50$&$\bm{12.47}$&$19.39$&$16.14$
\\
  & 5 & $13.21$&$19.60$&$12.49$&$19.57$&$16.22$\\
\midrule
\multirow{5}{*}{5$^{\ddagger}$} 
  & 1& $14.49$&$21.38$&$13.68$&$21.06$&$17.65$\\
  & 2& $13.40$&$19.90$&$12.75$&$19.64$&$16.42$
\\
  & 3& $13.20$&$19.70$&$12.62$&$19.59$&$16.28$
\\
  & 4& $13.06$&$19.23$&$12.24$&$19.31$&$15.96$
\\
  & 5$^{\ddagger}$& $\bm{12.64}$&$\bm{19.00}$&$\bm{12.19}$&$\bm{18.75}$&$\bm{15.65}$\\

\midrule
\multirow{5}{*}{6} 
  & 1& $14.53$&$21.29$&$13.86$&$21.03$&$17.68$\\
  & 2&  $13.57$&$20.06$&$12.85$&$19.82$&$16.58$\\
  & 3& $\bm{13.25}$&$\bm{19.51}$&$\bm{12.59}$&$\bm{19.55}$&$\bm{16.23}$\\
  & 4&$13.37$&$19.93$&$12.81$&$19.77$&$16.47$\\
  & 5&  $13.68$&$19.99$&$13.07$&$20.02$&$16.69$ \\


\midrule
\midrule
\multicolumn{7}{c}{Within-speaker} \\

\multicolumn{7}{l}{APC (\textit{training set: unlab-600})}\\

\multirow{5}{*}{3}
  & 1& $9.37$&$11.70$&$8.78$&$12.16$&$10.50$ \\
  & 2& $9.18$&$11.47$&$8.54$&$12.07$&$10.32$
\\
  & 3& $8.88$&$\bm{11.28}$&$8.41$&$11.60$&$10.04$
\\
  & 4& $8.89$&$11.34$&$8.49$&$11.51$&$10.06$
\\
  & 5& $\bm{8.76}$&$11.30$&$\bm{8.39}$&$\bm{11.43}$&$\bm{9.97}$\\
 \midrule
\multirow{5}{*}{4} 
  & 1 &$9.36$&$11.78$&$8.64$&$12.20$&$10.50$\\
  & 2& $9.22$&$11.62$&$8.64$&$12.15$&$10.41$
\\
  & 3& $8.82$&$11.22$&$8.26$&$11.50$&$9.95$
\\
  & 4& $8.81$&$\bm{11.18}$&$\bm{8.19}$&$\bm{11.34}$&$\bm{9.88}$
\\
  & 5& $\bm{8.63}$&$11.32$&$8.21$&$11.43$&$9.90$ \\

\midrule
\multirow{5}{*}{5$^{\ddagger}$} 
  & 1& $9.38$&$11.81$&$8.62$&$12.17$&$10.50$\\
  & 2 &$8.79$&$11.13$&$\bm{8.19}$&$11.39$&$\bm{9.88}$
\\
  & 3 &$\bm{8.69}$&$11.19$&$8.28$&$\bm{11.37}$&$\bm{9.88}$
\\
  & 4 &$8.78$&$11.23$&$8.32$&$11.42$&$9.94$
\\
  & 5$^{\ddagger}$ &$8.83$&$\bm{11.07}$&$8.36$&$11.48$&$9.94$\\
\midrule
\multirow{5}{*}{6} 
  & 1& $9.37$&$11.93$&$8.76$&$12.21$&$10.57$  \\
  & 2& $9.07$&$11.44$&$8.49$&$11.69$&$10.17$   \\
  & 3& $\bm{8.91}$&$\bm{11.36}$&$\bm{8.44}$&$\bm{11.60}$&$\bm{10.08}$ \\
  & 4& $9.07$&$11.47$&$8.56$&$11.63$&$10.18$\\
  & 5& $9.33$&$11.60$&$8.80$&$11.84$&$10.39 $\\
\midrule[0.2pt]
\bottomrule
\end{tabular}%

}
\label{tab:supple_exp_results_apc_paras}
\end{table}
\end{document}